\def\lesssim{\mathrel{\hbox{\rlap{\hbox{\lower4pt\hbox{$\sim$}}}\hbox{$<$}}}}  
  
\def\gtrsim{\mathrel{\hbox{\rlap{\hbox{\lower4pt\hbox{$\sim$}}}\hbox{$>$}}}}

\def\ll_lsun{log$({L/\rm L_{\odot}})$~}  
  
\def\masa_msun{$M/ \rm M_{\odot}$~}  
  
\def\m_mstar{$M/M_{*}$~}

\def\pg{\mbox{PG~1159$-$035}}  
\def\rx{\mbox{RX~J2117.1+3412}}  
\def\pp{\mbox{PG~0122+200}}

\documentclass{aa}  
\usepackage{graphicx}  
          
\begin{document}  
  
\title{Asteroseismological 
measurements on \pg, the prototype 
of the GW Vir variable stars}  
  
\author{A. H. C\'orsico$^{1,2}$\thanks{Member of the Carrera del Investigador  
Cient\'{\i}fico y Tecnol\'ogico, CONICET, Argentina.},  
L. G. Althaus$^{1,2\star}$,  S. O. Kepler$^3$, 
J. E. S. Costa$^3$, \and M. M. Miller 
Bertolami$^{1,2}$\thanks{Fellow of CONICET, Argentina.}}   
  
\offprints{A. H. C\'orsico}  
  
\institute{  
$^1$   Facultad   de   Ciencias  Astron\'omicas   y   Geof\'{\i}sicas,
Universidad  Nacional de  La Plata,  Paseo del  Bosque S/N,  (1900) La
Plata, Argentina.\\ $^2$ Instituto de Astrof\'{\i}sica La Plata, IALP,
CONICET-UNLP\\ $^3$ Instituto de F\'isica, Universidade Federal do Rio
Grande do Sul, 91501-970 Porto Alegre, RS, Brazil \\
\email{acorsico,althaus@fcaglp.unlp.edu.ar; kepler,costajes@if.ufrgs.br} }  
    
\date{Received; accepted}  
  
\abstract{}{An asteroseismological study of \pg, the prototype of the GW Vir 
variable stars, has been performed on the basis 
of detailed and full  PG1159 evolutionary models presented by
Miller  Bertolami   \&  Althaus  (2006).}{We   carried  out  extensive
computations  of  adiabatic   $g$-mode  pulsation  periods  on  PG1159
evolutionary models with stellar  masses spanning the range $0.530$ to
$0.741  M_{\odot}$.   These  models  were  derived  from  the  complete
evolution  of progenitor  stars, including  the thermally  pulsing AGB
phase and the born-again  episode.  We constrained the stellar  mass of \pg\
by comparing  the observed period  spacing with the  asymptotic period
 spacing and with the average  of the computed period spacings. We also
employed  the   individual observed  periods reported  by  Costa et
al. (2007)  to find a representative seismological  model for \pg.}{We
derive  a stellar  mass in  the range  $0.56-0.59 M_{\odot}$  from the
period-spacing data alone. We also  find, on the basis of a period-fit
procedure, an  asteroseismological model representative  of \pg\ that
reproduces the observed  period pattern with an average  of the period
differences of $\overline{\delta
\Pi_i}= 0.64 - 1.03$ s, consistent with the expected model uncertainties. 
The model has  an effective temperature $T_{\rm eff}=  
128\, 000^{+8\, 600}_{-2\, 600}$ K, a
stellar mass $M_*= 0.565^{+0.025}_{-0.009} M_{\odot}$, 
a surface gravity $\log g= 7.42^{+0.21}_{-0.12}$,
a stellar  luminosity and radius  of $\log(L_*/L_{\odot}) =  2.15\pm 0.08$ and
$\log(R_*/R_{\odot})=  -1.62^{+0.06}_{-0.09}$, and  a 
He-rich  envelope
thickness  of  $M_{\rm env}=  0.017  M_{\odot}$.  The  results of  the
period-fit analysis carried out in  this work suggest that the surface
gravity of \pg\ would be $1\,\sigma$ larger than the spectroscopically
inferred gravity. For our best-fit  model of \pg, all of the pulsation
modes are characterized  by positive rates of period  changes, at odds
with the measurements by Costa \& Kepler (2007).}{}
    
  
\authorrunning{C\'orsico et al.}  
  
\titlerunning{Asteroseismological measurements on \pg}  
  
\maketitle  
  
   
\section{Introduction}  
\label{intro}  
  
\pg\ (GW  Virginis) is  the prototype of  both the class  of pulsating
PG1159 stars ---commonly  known as GW Vir or  DOV variables--- and the
spectral  class  of  hot,  hydrogen-deficient,  post-asymptotic  giant
branch  (AGB) stars  with surface  layers rich  in helium,  carbon and
oxygen  ---the PG1159  stars  (see  Werner \&  Herwig  2006).  GW  Vir
variables exhibit  multiperiodic luminosity fluctuations  with periods
in the  range $5-50$ minutes,  attributable to nonradial  gravity mode
pulsations driven by the  $\kappa$-mechanism due to partial ionization
of  carbon and oxygen  in the  outer layers  (Starrfield et  al. 1983;
Gautschy   et  al.   2005;   C\'orsico  et   al.   2006;  Quirion   et
al. 2007).  PG1159 stars are considered the  evolutionary link between
post-AGB stars and most of  the hydrogen-deficient white dwarfs. It is
accepted that  these stars have  their origin in a  born-again episode
induced by a post-AGB helium thermal pulse (see Herwig 2001, Bl\"ocker
2001, Lawlor \& MacDonald 2003, Althaus et al.  2005, Miller Bertolami
et al. 2006 for recent references).

\begin{figure}  
\centering  
\includegraphics[clip,width=250pt]{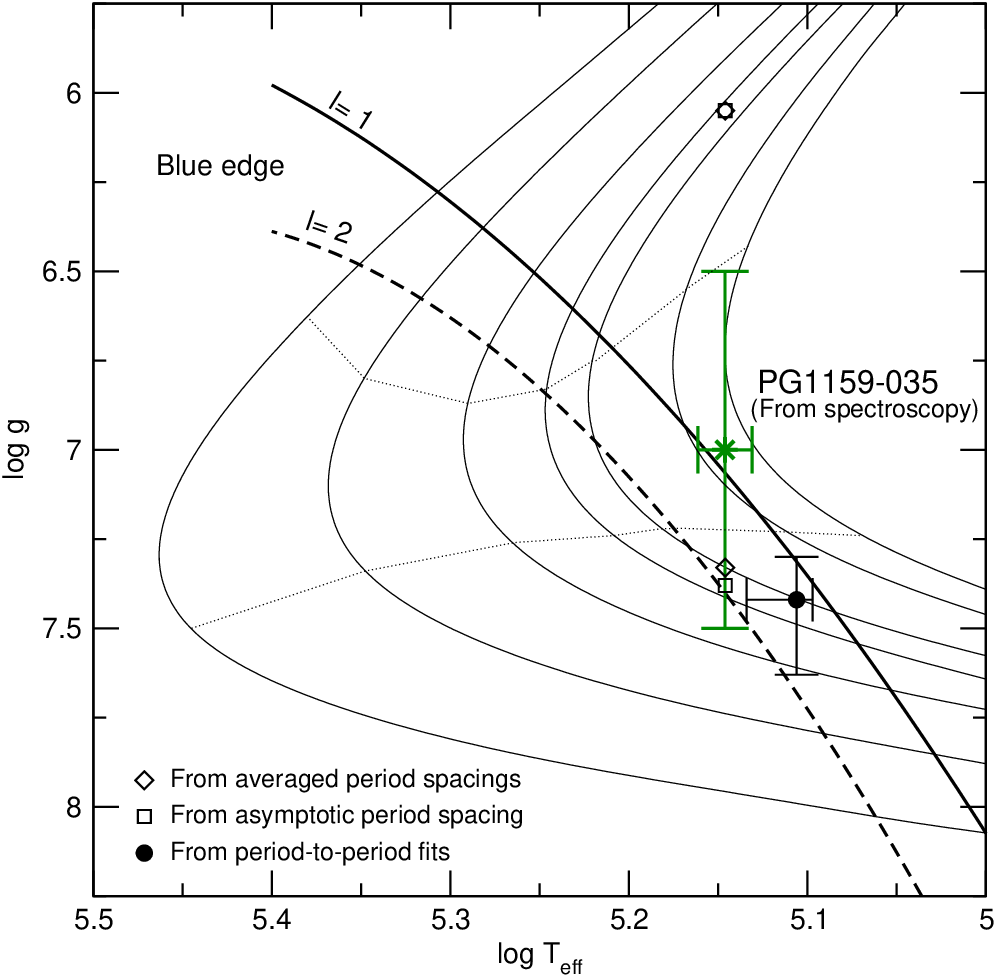}  
\caption{Our PG1159 full evolutionary  tracks in    
the $\log T_{\rm eff} -\log g$ plane, with stellar mass (from right to
left):  $0.530,  0.542,  0.565,   0.589,  0.609,  0.664$,  and  $0.742
M_{\odot}$.   The  location  of  \pg\  according  to  spectroscopy  is
indicated with  a star symbol.  The hollow  squares (diamonds) display
the  predictions  of the  asymptotic  (computed)  period spacings  (\S
\ref{sect-aps} and\S
\ref{sect-psp}). The filled  circle shows the location of
\pg\    according    to    the    period   fitting    procedure    (\S
\ref{fitting}). Thin dotted lines define three different regimes for 
the  theoretical  rate  of  period  change:  models  exhibit  positive
(negative)  rates of  period change  below (above)  the  lower (upper)
line, and positive  and negative rates (between both  lines).  The hot
boundaries  of the  theoretical  dipole ($\ell=  1$,  solid line)  and
quadrupole  ($\ell=  2$,  dashed   line)  GW  Vir  instability  domain
---according  to  C\'orsico et  al.   (2006)---  are  also shown.  For
details, see text.}
\label{teff-g}  
\end{figure}  

\pg\ is  characterized by $T_{\rm eff}=  140\, 000 \pm 5\,  000$ K and
$\log g= 7.0  \pm 0.5$ (Werner et al.  1991).   Since the discovery of
its photometric variations by McGraw et al.  (1979), the star has been
the focus of intensive  scrutiny.  In an impressive pulsational study,
Winget et al.  (1991) (WEA91) identified 122 peaks  in the periodogram
(Fourier  spectrum) of \pg\  obtained with  the Whole  Earth Telescope
(WET; Nather  et al. 1990).  The  peaks were attributed  to high order
nonradial  $g$-modes with  periods between  $\approx 300-1000$  s, and
with period spacings of $\Delta \Pi_{\ell= 1}^{\rm O}= 21.50 \pm 0.03$
s and $\Delta \Pi_{\ell= 2}^{\rm O}= 12.67 \pm 0.03$ s.  The mode with
the largest amplitude (7.2 mma),  which was identified as a $\ell= 1$,
$m= +1$, has a period of $\approx  516$ s.  By using a larger data set
from different years  (1983, 1985, 1989, 1993, and  2002) and improved
data  reduction  and  data  analysis,  Costa et  al.   (2007)  (CEA07)
identified 76  additional pulsation modes, enlarging to  198 the total
number of pulsation  periods for \pg, and placing it  as the star with
the  largest number  of  modes  detected after  the  Sun. CEA07  found
$\Delta  \Pi_{\ell=  1}^{\rm  O}=   21.43  \pm  0.03$  s  and  $\Delta
\Pi_{\ell= 2}^{\rm O}= 12.38 \pm 0.01$ s.

The determination of the stellar mass  of \pg\ has been the subject of
numerous  investigations.   The stellar  mass  of pulsating  pre-white
dwarfs    can   be    constrained    from   asteroseismology    ---the
asteroseismological mass--- either through the observed period spacing
(see,  for instance,  Kawaler  \& Bradley  1994  (KB94); C\'orsico  \&
Althaus 2006)  or by  means of the  individual observed  periods (see,
e.g., KB94, C\'orsico \& Althaus  2006, C\'orsico et al. 2007ac).  The
study of WEA91 found a mass  of $0.586\pm 0.003 M_{\odot}$ for \pg\ on
the basis of  the observed period spacing for $\ell=  1$ and $\ell= 2$
modes and an asymptotic analysis  based on the PG1159 structure models
of Kawaler (1986, 1987, 1988).  From a detailed period fit, KB94 found
$M_*= 0.59 \pm  0.01 M_{\odot}$ for \pg.  C\'orsico  \& Althaus (2006)
found  a  stellar mass  of  $0.56  M_{\odot}$  from a  period  fitting
procedure  based on PG  1159 evolutionary  models with  several masses
created artificially  from the full  sequence of $0.589  M_{\odot}$ of
Althaus et al. (2005).  Finally,  with the enlarged set of periods for
\pg, CEA07 obtain  a value of $M_*= 0.59 \pm  0.02 M_{\odot}$ by using
the parameterization of the asymptotic period spacing of KB94.
  
The total  mass of  PG1159 stars can  also be estimated  through the
comparison of the  spectroscopic values of $T_{\rm eff}$  and $g$ with
evolutionary tracks  ---the spectroscopic mass.   On the basis  of the
evolutionary tracks  of O'Brien \&  Kawaler (2000), Dreizler  \& Heber
(1998) derived a  stellar mass  of $0.54 \pm  0.1 M_{\odot}$  for \pg.
The most recent  determination is that of Miller  Bertolami \& Althaus
(2006), who also derived a stellar mass of $0.54 \pm 0.1 M_{\odot}$ 
on the basis of PG1159  evolutionary   models with different stellar 
masses that   take  fully  into   
account  the evolutionary history of their progenitor stars, 
particularly the thermally pulsing and born again phases. 

The discrepancy between  the asteroseismological and the spectroscopic
mass of \pg\ has been partially alleviated by the employment of Miller
Bertolami  \&  Althaus  (2006)   PG1159  evolutionary  models 
and the average of the computed period spacings  in  the
determination  of  the  asteroseismological  mass.  In  fact,  at  the
effective temperature  of \pg, the  average of the computed  $\ell= 1$
period spacings  is consistent with  a stellar mass of  $\approx 0.558
M_{\odot}$ (C\'orsico et al. 2006).

The aim  of this  paper is to  present a  detailed asteroseismological
study of \pg\ based on the evolutionary models of Miller Bertolami
\&  Althaus  (2006)  and  the observational  data  of  CEA07.  
The  present study  is  the  third time  that  such consistent  PG1159
evolutionary  models are  employed  in individual  asteroseismological
modeling of  pulsating PG1159 stars,  the first  application being 
the study carried out by C\'orsico et  al. (2007a) for the hottest known 
GW Vir star \rx, and the second   being the study  
performed by C\'orsico et
al.  (2007c)  for the  coolest member of  the class, \pp.  
  
The  paper is organized  as follows:  in the  next Section  we briefly
describe      our       PG1159      evolutionary      models.       In
Sect. \ref{period-spacing} we derive the stellar mass of \pg\ by means
of  the observed period  spacing.  In  Sect.  \ref{fitting}  we derive
structural  parameters  of  this  star  by  employing  the  individual
observed periods.   In this  section we derive  an asteroseismological
model representative of \pg\ (\S \ref{best-fit}), and discuss aspects 
such as the mode-trapping properties (\S \ref{mode-trapping}) and 
the rates of period changes (\S \ref{period-changes}) 
of the best-fit model, and the asteroseismological distance of \pg\ 
(\S \ref{distance}). Finally, in Sect.
\ref{conclusions}  we  summarize  our   main  results  and  make  some  
concluding remarks.
  
\begin{figure*}  
\centering  
\includegraphics[clip,width=400pt]{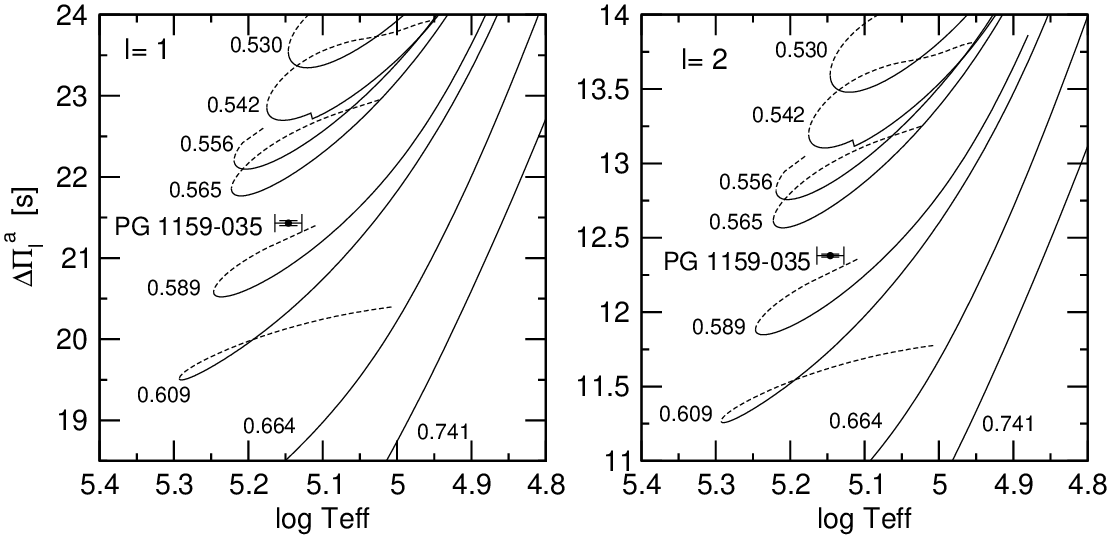}  
\caption{The dipole (left) and quadrupole (right) asymptotic period spacing   
($\Delta \Pi^{\rm a}_{\ell}$) for different stellar masses in terms of
the effective temperature.  The stages before (after) the evolutionary
``knee'' are depicted with  dashed (solid) lines.  Numbers along each
curve denote the stellar masses (in solar units).  The plot also shows
the location of PG 1159$-$035 according to CEA07.}
\label{aps}  
\end{figure*}  
  
\section{Evolutionary models and numerical tools}  
\label{evolutionary}  
  
The  pulsation  analysis  presented  in  this work  relies  on  a  new
generation  of stellar  models  that take  into  account the  complete
evolution of  PG1159 progenitor stars. Specifically, the stellar models  
were extracted from the evolutionary
calculations  recently  presented by  Althaus  et  al. (2005),  Miller
Bertolami \& Althaus (2006), and C\'orsico et al. (2006), who computed
the complete evolution of model  star sequences with initial masses on
the  ZAMS in  the range  $1 -  3.75 M_{\odot}$.   All of  the post-AGB
evolutionary  sequences  computed with the LPCODE evolutionary code 
(Althaus et al. 2005) have  been  followed through  the  very  late
thermal pulse  (VLTP) and the  resulting born-again episode  that give
rise to the H-deficient, He-, C- and O-rich composition characteristic
of  PG1159 stars.  The masses  of  the resulting  remnants are  0.530,
0.542, 0.556, 0.565, 0.589,  0.609, 0.664, and $0.741 M_{\odot}$.  The
evolutionary tracks in  the $\log T_{\rm eff} - \log  g$ plane for the
PG1159 regime are displayed in Fig. \ref{teff-g}.
  
The use  of these evolutionary tracks constitutes  an improvement
with respect  to previous asteroseismological  studies.  
These evolutionary  sequences  reproduce  (1)  the  spread  in
surface chemical  composition observed in PG1159 stars,  (2) the short
born-again times of V4334 Sgr (Miller Bertolami et al. 2006 and Miller
Betolami \& Althaus 2007a),  (3) the location  of the GW  Vir instability
strip  in  the  $\log  T_{\rm  eff}  - \log  g$  plane  (C\'orsico  et
al. 2006),  (4) the expansion  age of the planetary  nebula of 
\rx\ (C\'orsico  et al. 2007ab), and (5) the period spectrum of
the coolest GW Vir star, \pp\ (C\'orsico et al. 2007c). 
  
We  computed $\ell=  1$  and $\ell=  2$  $g$-mode adiabatic  pulsation
periods with the same  numerical code
and  methods we  employed  in  our previous  works  (see C\'orsico  \&
Althaus  2006 for  details).   We analyzed  about  3000 PG1159  models
covering a wide range of effective temperatures ($5.4 \gtrsim
\log(T_{\rm eff}) \gtrsim 4.8$) and luminosities ($0 \lesssim  
\log(L_*/L_{\odot}) \lesssim 4.2$), and a range of stellar masses  
($0.530 \leq M_*/M_{\odot} \leq 0.741$).
  
\begin{figure*}  
\centering  
\includegraphics[clip,width=400pt]{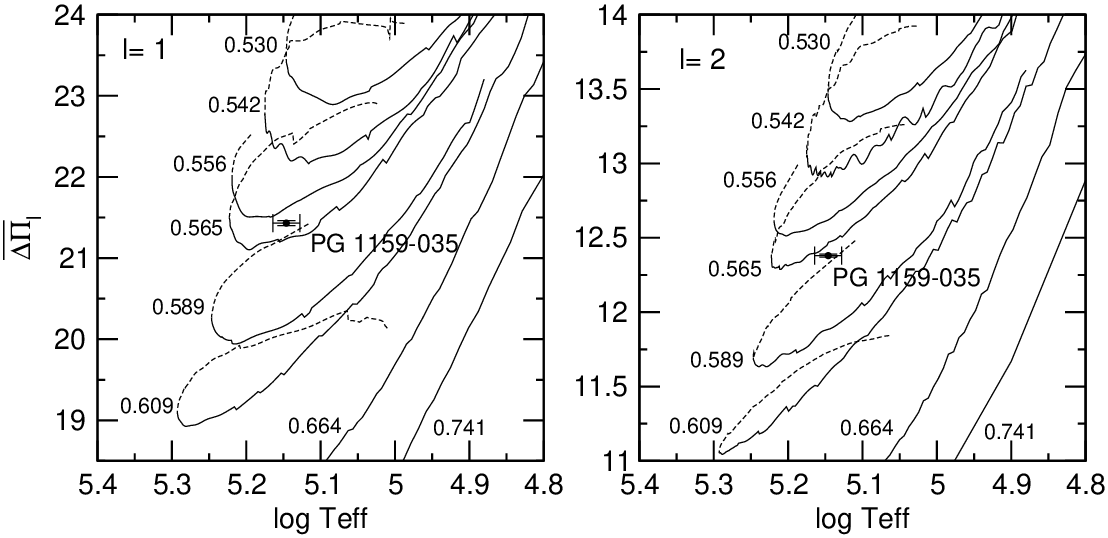}  
\caption{Same as Fig. \ref{aps}, but for the average   
of the computed period spacings ($\overline{\Delta \Pi_{\ell}}$).}  
\label{psp}  
\end{figure*}  
  
\section{Mass determination from the observed period spacing}  
\label{period-spacing}  
  
In this section we constrain the stellar mass of \pg\ by comparing the
asymptotic  period spacing  and  the average  of  the computed  period
spacings with the {\it observed} period spacing (\S \ref{sect-aps} and
\ref{sect-psp}, respectively). These methods take full 
advantage  of the  fact that  the period  spacing of  PG1159 pulsators
depends primarily  on the stellar  mass, and weakly on  the luminosity
and  the  He-rich envelope  mass  fraction  (KB94; C\'orsico \& Althaus 2006).

Most of the published asteroseismological studies on PG1159 stars rely
on the asymptotic  period spacing to infer the total  mass of GW Vir
pulsators, the notable exception being the works by KB94 
for  \pg, C\'orsico  et  al. (2007a)  for  \rx\  and
C\'orsico et al. (2007c) for \pp.

The precise evolutionary  status of \pg\ is not  known {\it a priori}.
Thus, we  assume two  possible cases in our  following analysis:
one  situation in which  the star  is still  on the  rapid contraction
phase before  reaching its maximum effective  temperature (i.e. before
the evolutionary ``knee'')  and the other situation in  which the star
has just settled onto the white dwarf  cooling track (i.e.  after 
the evolutionary ``knee'').

\subsection{First method: comparison with the asymptotic period 
spacing, $\Delta  \Pi_{\ell}^{\rm  a}$}  
\label{sect-aps}

Fig.  \ref{aps}  displays the asymptotic period spacing  for $\ell= 1$
(left)  and $\ell= 2$  (right) modes  as a  function of  the effective
temperature  for  different  stellar  masses.   Also  shown  in  these
diagrams is  the location  of \pg,  with $T_{\rm eff}=  140 \pm  5$ kK
(Werner et  al.  1991), and  $\Delta \Pi^{\rm O}_{\ell= 1}=  21.43 \pm
0.03$ s and $\Delta \Pi^{\rm O}_{\ell= 2}= 12.38 \pm 0.01$ s (CEA07). 
The asymptotic period spacing is computed as 
$\Delta \Pi_{\ell}^{\rm a}=
\Pi_0 / \sqrt{\ell(\ell+1)}$, where $\Pi_0= 2 \pi^2 [ \int_{r_1}^{r_2}  
(N/r)   dr]^{-1}$,   and   $N$   the   Brunt-V\"ais\"al\"a   frequency
(e.g. Tassoul et al. 1990).
  
From the  comparison between the observed  $\Delta \Pi^{\rm O}_{\ell}$
and $\Delta \Pi_{\ell}^{\rm a}$  we found a stellar mass between 
$\approx  0.585  M_{\odot}$  (if   the  star  is  located  before  the
evolutionary  knee) and  $\approx  0.577 M_{\odot}$  (if  the star  is
located  after  the evolutionary  knee),  irrespective  of the  $\ell$
value;  see Table  \ref{spacing}.  The  inferred range  of mass  is in
agreement  with the  value  $M_* \approx  0.59  M_{\odot}$ derived  by
WEA91 and KB94 --- and also in agreement with the value
derived in CEA07 from the KB94 parameterization --- on
the basis of an asymptotic analysis of the period spacing.

As in our  previous works, we must emphasize  that the derivation of
the  stellar mass  using  the  asymptotic period  spacing  may not  be
entirely  reliable in  pulsating PG1159  stars (see Althaus et al. 2007).  
This  is  because the asymptotic predictions are strictly valid in the 
limit of very high radial order (long periods) and for chemically homogeneous
stellar  models, while  PG1159  stars are  supposed  to be  chemically
stratified  and characterized  by strong  chemical gradients  built up
during the progenitor  star life.  A more realistic  approach to infer
the stellar  mass of  PG1159 stars  is to compare  the average  of the
computed period spacings with the observed period spacing.

\begin{table}
\centering
\caption{Stellar mass (in $M_{\odot}$) of \pg\ derived 
from the period-spacing data.}
\begin{tabular}{c|cc|cc}
\hline
\hline
 & From & $\Delta \Pi^{\rm a}_{\ell}$ & 
 From & $\overline{\Delta \Pi_{\ell}}$ \\
 \cline{2-5}
& $\ell= 1$ & $\ell= 2$ & $\ell= 1$ & $\ell= 2$ \\
\hline
Before knee  & 0.585 & 0.585 & 0.586 & 0.587 \\
After knee   & 0.577 & 0.577 & 0.561 & 0.569 \\
\hline
\end{tabular}
\label{spacing}
\end{table} 

\subsection{Second method: comparison with the average of the computed 
period spacings, $\overline{\Delta \Pi_{\ell}}$}  
\label{sect-psp}

The quantity  $\overline{\Delta \Pi_{\ell}}$ is  assessed by averaging
the computed forward period spacings ($\Delta \Pi_{k}= \Pi_{k+1}-
\Pi_{k}$) in  the range of the  observed periods in  \pg: $390-990$ s
for $\ell=  1$ and $350-860$ s  for $\ell= 2$  (see Tables 4 and  12 of
CEA07).

In Fig.  \ref{psp}  we show the run of average  of the computed period
spacings for  $\ell= 1$ (left) and  for $\ell= 2$ (right)  in terms of
the  effective   temperature  for  all  of   our  PG1159  evolutionary
sequences. Note that the run of $\overline{\Delta \Pi_{\ell}}$ depends
on the  range of periods on  which the average of  the computed period
spacing  is done. Again,  the stages  before (after)  the evolutionary
knee  are  depicted  with  dashed  (solid) lines.   By  adopting  the
effective temperature of \pg\  as given by spectroscopy ($T_{\rm eff}=
140  \pm  5$  kK) we  found  a  stellar  mass  in the  range  $\approx
0.586-0.587 M_{\odot}$ (if the star is located before the evolutionary
knee) and $\approx 0.56-0.57 M_{\odot}$  (if the star is located after
the  evolutionary knee);  see  Table \ref{spacing}.   Note that  these
values are somewhat different to  the mass derived in C\'orsico et al.
(2006) ($\approx 0.558  M_{\odot}$) because in that  paper the authors
used a different range of periods to compute the average of the period
spacing, and a older value for observed period spacing value.

From Table \ref{spacing}  we see that, if the  star is evolving before
the knee, the value of $M_*$ derived from the $\overline{\Delta
\Pi_{\ell}}$ is only slightly larger ($0.25 \%$) than the value inferred 
from $\Delta  \Pi_{\ell}^{\rm a}$. If  star is evolving  after the
knee, instead, the stellar mass derived from $\overline{\Delta
\Pi_{\ell}}$ is $\sim 3 \%$ lower than the mass derived from $\Delta
\Pi_{\ell}^{\rm a}$. 

In the plot  of Fig.  \ref{teff-g}, the squares correspond
to the approximate location of the star as predicted by $\Delta
\Pi^{\rm a}_{\ell}$, whereas the diamonds depict the
situation as predicted by $\overline{\Delta \Pi_{\ell}}$. For the case
in which  the star is evolving  before the knee,  the location derived
from $\Delta  \Pi^{\rm a}_{\ell}$  coincides with that  predicted from
$\overline{\Delta  \Pi_{\ell}}$.   In  this  case the  star  would  be
located well inside the theoretical  GW Vir instability strip, but the
surface  gravity would  be excessively  low  ($\log g  \approx 6$)  as
compared  with the spectroscopically  inferred value  ($\log g=  7 \pm
0.5$). Thus, we will discard this solution.  On the other hand, if the
star is evolving after the knee, the surface gravity would be somewhat
larger  ($\log  g \approx  7.5$)  but  well  within the  spectroscopic
uncertainty,  although   in  this  case  the   nonadiabatic  model  is
pulsationally  stable  against  $\ell=  1$  modes,  contradicting  the
observational evidence.

\section{Constraints from the individual observed periods}  
\label{fitting}  

In this approach we seek a pulsation model that best matches the
\emph{individual}  pulsation  periods of  \pg.   
The goodness  of the match  between the theoretical  pulsation periods
($\Pi_k$)  and the  observed individual  periods ($\Pi_i^{\rm  O}$) is
measured by means of a merit function defined as 

\begin{equation}
\chi^2(M_*, T_{\rm
eff})= \frac{1}{n}\sum_{i=1}^{n} \min[(\Pi_i^{\rm  O}- \Pi_k)^2], 
\label{chi2}
\end{equation}

\noindent where $n$ is
the  total number  of the  observed periods  considered. Note 
that taking the square root of $\chi^2$ we obtain the standard 
deviation between the observed and the theoretical periods. 
The  PG 1159 model that shows  the lowest value of $\chi^2$ will  
be adopted as the ``best-fit model''.
This  approach ---which is usually called the
forward method in asteroseismology--- has also been used in the context of 
pulsating PG1159 stars by C\'orsico \&
Althaus (2006) and C\'orsico et al. (2007ac). We evaluate the function
$\chi^2(M_*, T_{\rm eff})$ for stellar masses of $0.530, 0.542, 0.556,
0.565, 0.589, 0.609, 0.664$, and $0.741 M_{\odot}$.  For the effective
temperature we employed a finer grid  ($\Delta T_{\rm eff}\sim 100-1000$ K)  
which  is given  by  the time  step  adopted  in our  evolutionary 
calculations.

\begin{figure*} 
\centering 
\includegraphics[clip,width=500pt]{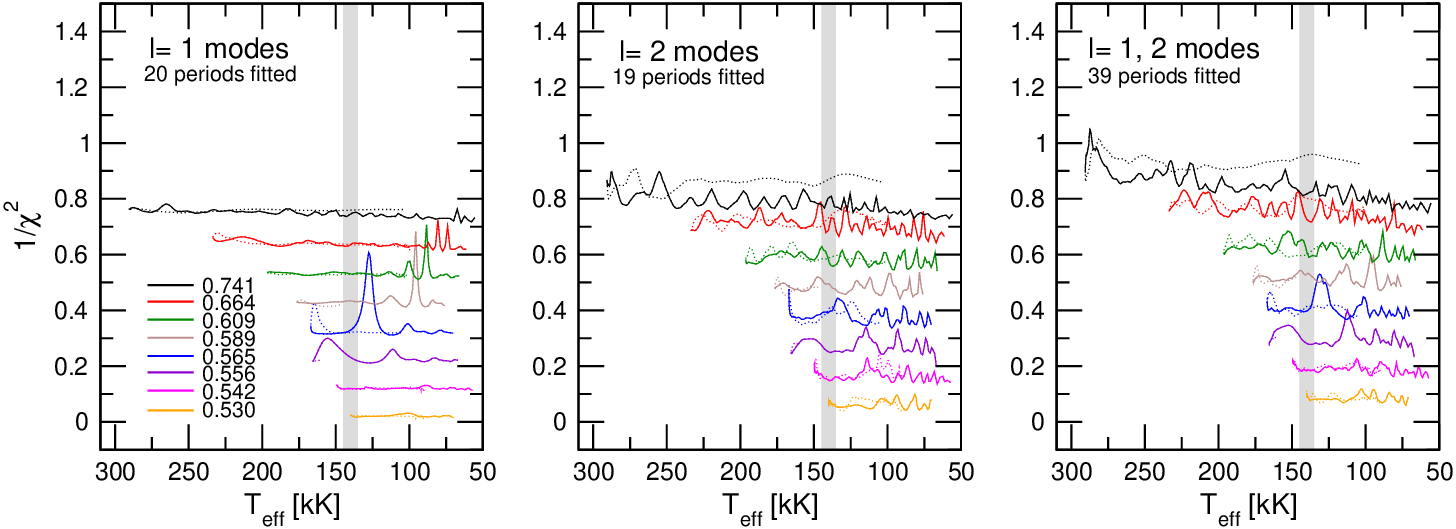} 
\caption{The inverse of the quality function corresponding to 
period fits considering  $\ell= 1$ modes only (left  panel), $\ell= 2$
modes  only (centre  panel),  and $\ell=  1,  2$ modes  simultaneously
(right panel).   Lines with different colours  correspond to sequences
with  different stellar  masses.  Dotted (solid)  lines correspond  to
stages  before  (after)  the  maximum effective  temperature  of  each
evolutionary sequence is  reached. All of the observed  $m= 0$ periods
were extracted from Table 3 of WEA91.  For clarity, the
curves have  been artificially  shifted upward (with  a step  of 0.1),
except  for the lower-most curve which corresponds to the 
$0.530  M_{\odot}$ sequence.   The vertical  grey band
displays  the   \pg\  spectroscopic  effective   temperature  and  its
uncertainties. [Color
figure only available in the electronic version of the article] }
\label{todos}  
\end{figure*}  

\begin{figure*}  
\centering  
\includegraphics[clip,width=500pt]{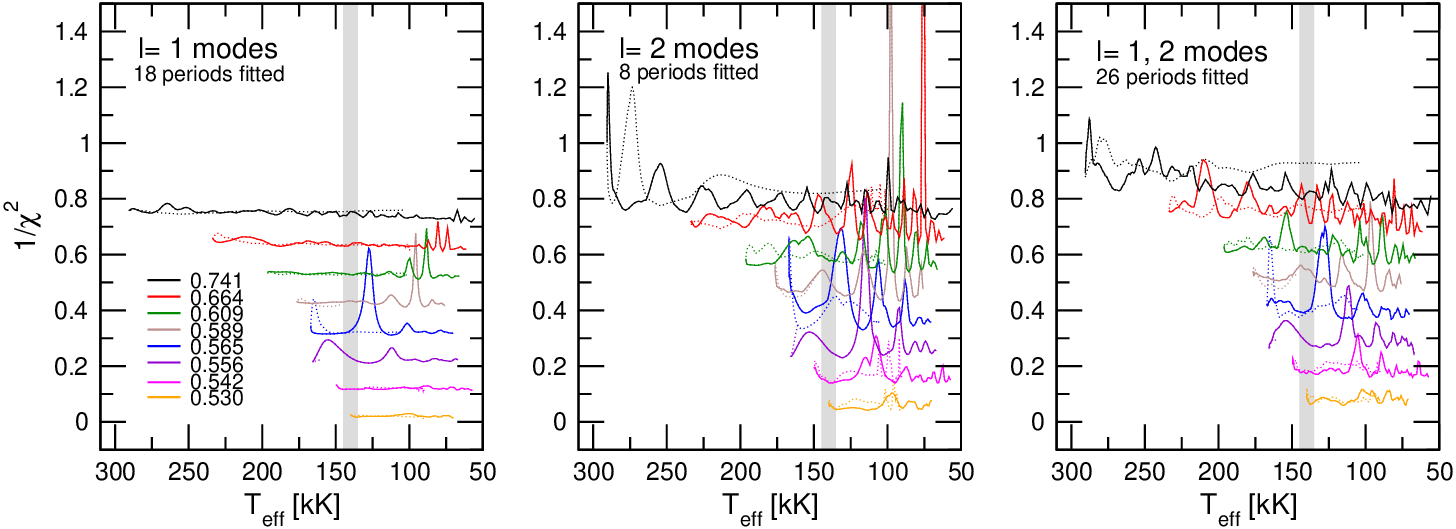}  
\caption{Same as Fig. \ref{todos}, but for the case of the observed 
$m= 0$ periods with identification secure only, extracted from Table 3
of WEA91. [Color
figure only available in the electronic version of the article]}
\label{seguros}  
\end{figure*}  
 
\subsection{The WEA91 observed periods}  
\label{winget}  
  
\begin{table}  
\centering  
\caption{Observed periods ($\Pi_i^{\rm O}$) of identified 
$m=  0$  and  $\ell=1,  2$  modes  for \pg.}
\begin{tabular}{cc|cc|cc|cc}  
\hline  
\hline  
WEA91     & & CEA07 & & WEA91 & & CEA07 & \\
$\ell= 1$ &  & $\ell= 1$ & &  $\ell= 2$ & & $\ell= 2$& \\  
\hline  
       &   & 390.30 & s & 339.24  & s &         &   \\ 
       &   & 412.01 & s & 351.01  & s & 350.75  &   \\
430.04 & s & 432.37 &   & 361.76  & s & 363.39  &   \\
450.71 &   & 452.06 & s &         &   & 376.03  &   \\ 
469.57 & s & 472.08 &   & 388.07  & s & 387.47  &   \\ 
494.85 & s & 494.85 & s & 398.89  &   & 400.06  & s \\ 
517.18 & s & 517.16 & s & 413.30  & s & 413.14  &   \\
538.16 & s & 538.14 & s & 425.03  & s & 425.04  &   \\
558.44 & s & 558.14 & s & 438.00  & s & (437.91)&   \\
581.29 &   & 579.12 &   &(453.62) &   & 449.43  &   \\
603.04 & s & 603.04 &   &         &   & 498.73  & s \\
622.60 & s & 622.00 &   &         &   & 511.98  & s \\
643.41 & s & 643.31 & s &         &   & 524.03  &   \\
666.22 & s & 668.09 &   &         &   & 536.37  &   \\
687.71 & s & 687.74 & s &         &   & 547.00  &   \\
707.92 & s & 709.05 &   &         &   & 561.99  & s \\
729.50 & s & 729.51 & s & 577.17  & s & 573.69  & s \\
753.12 & s & 752.94 &   &         &   & 585.26  &   \\
773.77 & s & 773.74 &   & 635.92  &   &         &   \\
790.94 & s & 791.80 &   & 649.67  &   &         &   \\
817.12 & s & 814.58 &   &         &   & 660.46  &   \\
840.02 & s & 838.62 & s &         &   & 672.20  &   \\
       &   & 861.72 &   &         &   & 684.48  &   \\
       &   & 883.67 &   & 694.88  &   & 696.83  &   \\
       &   & 903.19 & s &         &   & 709.87  &   \\
       &   & 925.31 & s &(734.87) &   & 746.38  & s \\     
       &   & 945.01 &   & 779.48  &   & 783.19  &   \\
       &   & 966.98 & s & 812.57  &   & 820.90  & s \\
       &   & 988.13 &   & 833.31  &   &         &   \\
       &   &        &   &         &   & 858.84  &   \\
       &   &        &   &(929.38) &   &         &   \\
       &   &        &   & 982.22  &   &         &   \\ 
\hline  
 20       & 18 & 29      & 14 &  19     & 8  & 26 & 7\\ 
\hline  
\hline  
\end{tabular}  
\label{periods}  
{\footnotesize  Note: parenthesis indicate $m= 0$ periods which are actually 
absent from the power spectrum. Their values are estimated by averaging 
the components $m= \pm 1$.}  
\end{table}  

We first performed period-to-period fits by considering the old 
set of \pg\ observed $m=0$ periods of  WEA91. 
They  are reproduced in the first and third columns of 
our Table \ref{periods} in units of seconds.  
A letter  ``s'' indicates a mode  with a secure
identification of $m$ and $\ell$. The last  row shows  the total  
number of  periods  of the corresponding column.
The results of our period fits are displayed in  
Figs. \ref{todos} and \ref{seguros},  where the quantity  
$(\chi^2)^{-1}$ in
terms of  $T_{\rm eff}$ for different stellar  masses is shown. 
In the interest of clarity, the curves are shifted upwards with a step of
0.1  starting from the  curve corresponding  to the  $0.530 M_{\odot}$
sequence.    The   spectroscopic   $T_{\rm eff}$ and   its
uncertainties  are depicted  with a  vertical grey  strip.  A  peak 
in the inverse of the merit function with  
$(\chi^2)^{-1} \gtrsim 0.3$ will be considered as a 
good match between the  theoretical and the observed periods. 

We started our analysis by performing a period fit considering the
supposed $m= 0,\ \ell= 1$ modes only  (first column of Table \ref{periods}).  
In this way, the value of $\ell$ of the computed periods 
is fixed to be $\ell= 1$ in our period fit procedure. 
The function  $(\chi^2)^{-1}$   is characterized by a rather smooth 
behaviour except for a few pronounced peaks at  stages after the turning  
point in $T_{\rm  eff}$ is reached (left panel  of Fig. \ref{todos}).  
In particular,  there exists a
primary  peak  that is  seen  for the  first  time  at high  effective
temperatures  ($\approx  160\,  000$  K)  for  the  $0.556  M_{\odot}$
sequence.  The peak  gradually shifts to lower $T_{\rm  eff}$ for the
sequences with higher  masses, lying at $\sim 65\, 000$  K for $M_* =
0.741 M_{\odot}$.  The  maximum amplitude of the peak  (i.e., the best
period match)  is reached for  a $0.565 M_{\odot}$ model  with $T_{\rm
eff} \sim  128\,000$ K, somewhat lower than  the spectroscopic 
effective temperature
of \pg\ ($T_{\rm eff}= 140\,000 \pm 5\, 000$ K).  In other words, 
if the $\ell= 1$ modes  were the  only  modes present in  \pg,  
our best  
asteroseismological solution  derived from a  period fit  should be  
a model  with $T_{\rm eff}=  127\,680$ K  and $M_*=  0.565  M_{\odot}$. 
The  quality of  our period  fit  is measured by  the  average  of the  
absolute  period differences,    $\overline{\delta   \Pi_i}=    
(\sum_{i=1}^n   |\delta \Pi_i|)/n$,  where $\delta \Pi_i=  
\Pi_i^{\rm O}  -\Pi_k$, and  by the root-mean-square residual, 
$\sigma_{_{\delta \Pi_i}}= \sqrt{(\sum |\delta \Pi_i|^2)/n} $.  
For this solution we obtain $\overline{\delta \Pi_i}= 1.32$ s and 
$\sigma_{_{\delta \Pi_i}}= 1.80$ s.

Next, we  carried out  a period  fit  taking into  account the  supposed
$m= 0,\ \ell= 2$ modes only (third column of Table \ref{periods}).
The  value of $\ell$ of the theoretical  periods is fixed to 
be $\ell= 2$. The results are depicted  in the centre
panel of Fig. \ref{todos}.  The behaviour of $(\chi^2)^{-1}$ is very 
different  as compared  with the  $\ell= 1$  period  fit.  Remarkably,
$(\chi^2)^{-1}$ exhibits  numerous peaks of  almost similar amplitude,
irrespective  of the  stellar mass.   This means  that  the quadrupole
period spectrum  of the star  could not be  fitted by a  unique PG1159
model, due to the existence of numerous and almost equivalent possible
solutions \footnote{This effect  can  be  understood on  the  basis that  the
pulsation  periods of  a specific  model that  evolves  towards higher
(lower)  effective  temperatures  generally decrease  (increase)  with
time.  If at  a given $T_{\rm eff}$ the computed  periods of the model
show  a  close  fit  to   the  observed  periods,  then  the  function
$(\chi^2)^{-1}$  reaches a local  maximum.  Later  when the  model has
evolved enough (heated or cooled), it is possible that the accumulated
period drift nearly matches  the period spacing between adjacent modes
($|\Delta k|= 1$).   In these circumstances, the periods  of the model
are able to  fit the periods of  the star again, as a  result of which
$(\chi^2)^{-1}$  shows other local maxima.}. We  conclude that  if the
$\ell=  2$  modes were  the  only  present in \pg,  a  satisfactory
asteroseismological solution could not be found.

Finally we made a period  fit  using the $m= 0,\ \ell=  1$ and  
$m= 0,\ \ell= 2$  modes simultaneously (first  and third columns of  
Table \ref{periods}). In this case the value of $\ell$ for the theoretical 
periods is not fixed but instead is obtained as an output of our 
period fit procedure, although the allowed values are 1 and 2.  
The  results  are  displayed  in  the  right  panel  of
Fig. \ref{todos}.  The $(\chi^2)^{-1}$ function shows the  same behaviour  
as in  the case  of the  $\ell= 2$  period fit.
Thus, we  are unable to  find an asteroseismological  model satisfying
simultaneously  both the  $\ell= 1$  and  $\ell= 2$  sets of  observed
periods.

\begin{figure*}  
\centering  
\includegraphics[clip,width=500pt]{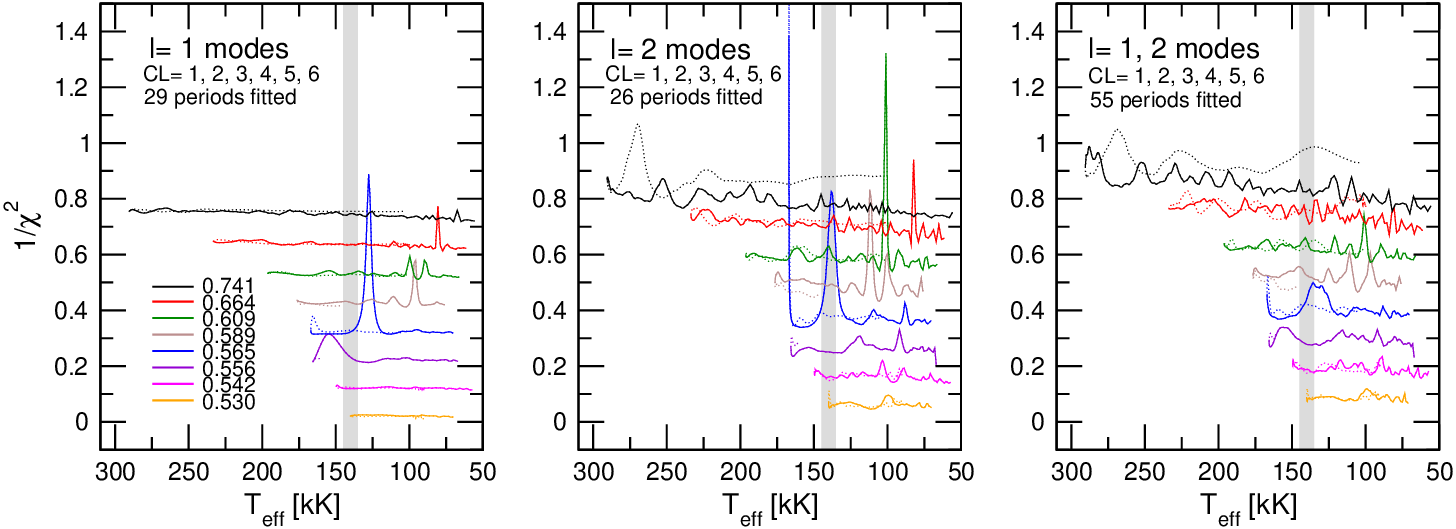}  
\caption{Same as Fig. \ref{todos}. Here, all of the observed 
$m=  0$ periods (having a Confidence Level of 1, 2, 3, 4, 5, or 6) have 
been extracted from Tables 4  and 12 of CEA07. 
[Color figure only available in the electronic version of the article]}
\label{cl123456}  
\end{figure*}  

\begin{figure*}  
\centering  
\includegraphics[clip,width=500pt]{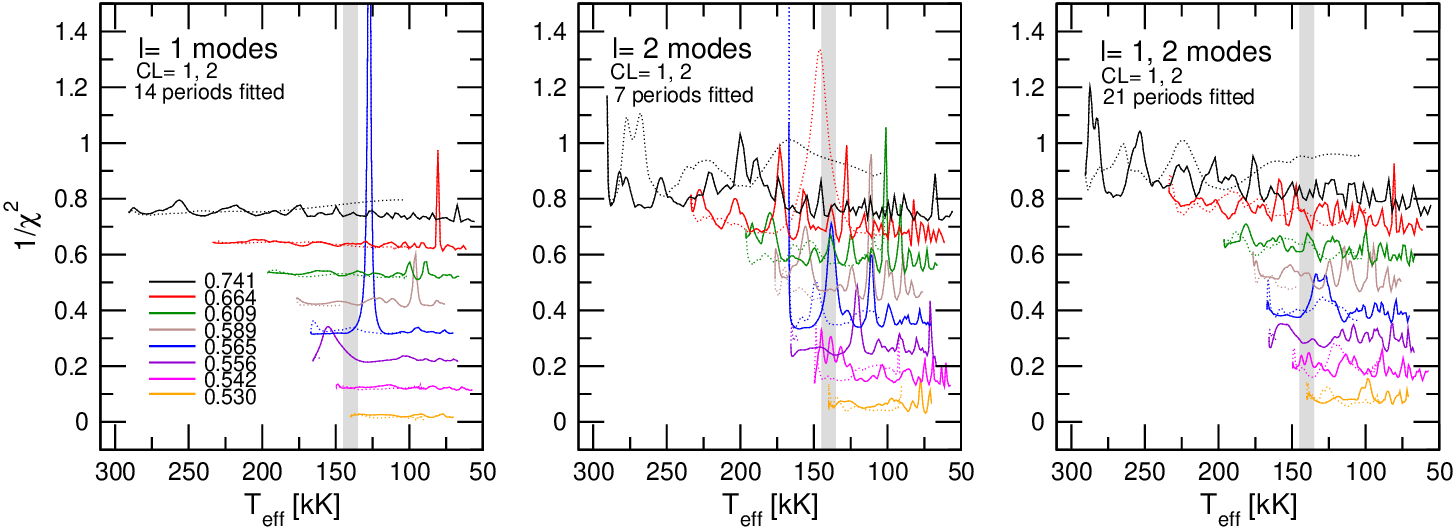}  
\caption{Same as Fig. \ref{cl123456}, but for the case of the observed 
$m= 0$  periods with Confidence Level  of 1 or 2  only, extracted from
Tables 4 and  12 of CEA07. [Color
figure only available in the electronic version of the article]}
\label{cl12}  
\end{figure*}  

Our next step  was to repeat the above period fits,  but this time by
considering only the  subset of observed periods for  which the $\ell$
and $m$ identification is considered as secure and free of ambiguities
by WEA91. The periods employed are labelled  with a  letter ``s'' in  
Table \ref{periods}.   As a result of this  selection, the number of 
observed  $\ell= 1$ periods is slightly reduced from 20 to 18,  
but for $\ell= 2$ modes the number is substantially reduced from 19 to  
8 periods.  

The results of the period fits are shown in Fig. \ref{seguros}. 
To begin with, for $\ell= 1$ modes (left panel) the behaviour of  
$(\chi^2)^{-1}$  remains  unaltered  as  compared with  the  situation
analyzed  in the left panel of Fig. \ref{todos}, due to both period  
fits being essentially the  same.  For the  best-fit model,
which is again the $0.565  M_{\odot}$ model at $T_{\rm eff}= 127\,680$
K, we  obtain $\overline{\delta \Pi_i}= 1.24$  s and $\sigma_{_{\delta
\Pi_i}}= 1.76$ s. For the $\ell=  2$ period fit, on the  other hand, 
the $(\chi^2)^{-1}$ function  experiences   appreciable  changes  as   
compared  with  the situation analyzed before.  
In  particular, the numerous peaks present in $(\chi^2)^{-1}$ 
exhibit larger amplitudes, revealing period matches
that are somewhat better as a  result of the smaller number of periods
to be fitted.

Finally,  the simultaneous  fit to  $\ell=  1$ and  $\ell= 2$  observed 
periods results  again in a $(\chi^2)^{-1}$ function with  a complex peaked
structure. Nevertheless, it is possible  to distinguish in this case a
prominent  peak corresponding to  the same  best-fit  model previously
found, with $T_{\rm eff}= 127\,680$  K and $M_*= 0.565 M_{\odot}$, and
the period fit in this case is characterized by $\overline{\delta
\Pi_i}= 1.25$ s and $\sigma_{_{\delta \Pi_i}}= 1.6$ s. A detailed 
comparison of  the observed periods  with the theoretical  periods for
this period fit  is provided in Table \ref{best-fit-a}.

Note that the
assignation of  the $\ell$ value  for the theoretical  periods differs
from the identification of the observed periods in some cases.

\subsection{The CEA07 observed periods}  
\label{costa}  

\begin{table}  
\centering  
\caption{Comparison between the observed $(\ell, m)= (1,0)$ 
and $(\ell, m)= (2,0)$
 periods ($\Pi_i^{\rm O}$, in units of seconds) for \pg\ taken  from WEA91  
and the theoretical  $(\ell, m)= (1,0)$ and $(\ell, m)= (2,0)$
periods ($\Pi_k$, in units of seconds) of the best-fit model 
with $M_*= 0.565 M_{\odot}$ and $T_{\rm eff}= 127\,680$  K. 
$\delta   \Pi_i= \Pi_i^{\rm O}- \Pi_k$ 
represents the period differences, $k$ the radial orders, and 
$\dot{\Pi}_k$ the rates of period change (in units of $10^{-11}$ s/s).} 
\begin{tabular}{cc|ccrccc}  
\hline  
\hline  
$\Pi_i^{\rm O}$ & $\ell^{\rm O}$ & $\Pi_k$ & $\ell$ &  
$\delta \Pi_i$ & $k$ &  $\dot{\Pi}_k$   \\
\hline  
339.24 & 2 & 340.79 & 2 & $-1.55$ & 25 & 0.88  \\
351.01 & 2 & 353.43 & 2 & $-2.42$ & 26 & 0.45  \\
361.76 & 2 & 365.04 & 2 & $-3.28$ & 27 & 0.85  \\
388.07 & 2 & 389.14 & 1 & $-1.70$ & 16 & 1.22  \\
413.30 & 2 & 412.06 & 1 & $ 1.24$ & 17 & 0.75  \\
425.03 & 2 & 426.63 & 2 & $-1.60$ & 32 & 0.67  \\  
430.04 & 1 & 431.47 & 1 & $-1.43$ & 18 & 0.54  \\  
438.00 & 2 & 439.28 & 2 & $-1.28$ & 33 & 0.97  \\  
469.57 & 1 & 465.22 & 2 & $ 4.35$ & 35 & 1.05  \\  
494.85 & 1 & 494.45 & 1 & $ 0.40$ & 21 & 0.70  \\  
517.18 & 1 & 516.41 & 1 & $ 0.77$ & 22 & 1.40  \\  
538.16 & 1 & 538.34 & 1 & $-0.18$ & 23 & 1.05  \\  
558.44 & 1 & 558.04 & 1 & $ 0.40$ & 24 & 0.77  \\  
577.17 & 2 & 576.23 & 2 & $ 0.94$ & 44 & 1.72  \\  
603.04 & 1 & 602.64 & 1 & $ 0.40$ & 26 & 1.03  \\  
622.60 & 1 & 621.89 & 1 & $ 0.71$ & 27 & 0.98  \\  
643.41 & 1 & 644.05 & 1 & $-0.64$ & 28 & 1.77  \\  
666.22 & 1 & 665.84 & 1 & $ 0.38$ & 29 & 0.93  \\  
687.71 & 1 & 686.62 & 1 & $ 1.09$ & 30 & 1.43  \\  
707.92 & 1 & 708.30 & 1 & $-0.38$ & 31 & 1.70  \\  
729.50 & 1 & 729.32 & 1 & $ 0.18$ & 32 & 1.48  \\  
753.12 & 1 & 754.38 & 2 & $-1.26$ & 58 & 1.50  \\  
773.77 & 1 & 772.65 & 1 & $ 1.12$ & 34 & 1.89  \\  
790.94 & 1 & 792.85 & 2 & $-1.91$ & 61 & 2.38  \\  
817.12 & 1 & 816.94 & 2 & $ 0.18$ & 63 & 1.73  \\  
840.02 & 1 & 837.38 & 1 & $ 2.64$ & 37 & 2.38  \\
\hline  
\hline  
\end{tabular}  
\label{best-fit-a}  
\end{table} 

\begin{table}  
\centering  
\caption{Observed $(\ell, m)= (1,0)$ periods 
($\Pi_i^{\rm O}$, in units of seconds) and amplitudes (in 
units of mma) for \pg\ taken  from CEA07, theoretical  $(\ell, m)= (1,0)$  
periods ($\Pi_k$ in units of seconds) of the best-fit model with 
$M_*= 0.565 M_{\odot}$ and $T_{\rm eff}= 127\,680$  K. 
$\delta   \Pi_i= \Pi_i^{\rm O}- \Pi_k$ represents 
the period   differences, $k$ are the radial orders, 
$\dot{\Pi}_k$ are the rates of period change (in units of $10^{-11}$ s/s), 
and $\ddot{\Pi}_k$ are the second order rates of period 
change (in units of $10^{-23}$ s/s$^2$).}
\begin{tabular}{lc|crccr}  
\hline  
\hline  
$\Pi_i^{\rm O}$ & $A_i^{\rm O}$ & $\Pi_k$ & $\delta \Pi_i$ & $k$ 
&  $\dot{\Pi}_k$      
& $\ddot{\Pi}_k$ \\
\hline  
390.30 s & 1.0 & 389.14 & $ 1.16$ & 16 & 1.22  & $-0.41$ \\
412.01 s & 0.6 & 412.06 & $-0.05$ & 17 & 0.76  & $-2.54$ \\
432.37   & 0.5 & 431.47 & $ 0.90$ & 18 & 0.54  & $ 1.54$ \\
452.06 s & 3.0 & 453.34 & $-1.28$ & 19 & 1.21  & $-0.26$ \\      
472.08   & 0.4 & 474.47 & $-2.39$ & 20 & 0.71  & $-2.30$ \\
494.85 s & 0.7 & 494.45 & $ 0.40$ & 21 & 0.70  & $ 2.08$ \\
517.16 s & 4.2 & 516.41 & $ 0.75$ & 22 & 1.40  & $ 0.14$ \\
538.14 s & 0.6 & 538.34 & $-0.20$ & 23 & 1.05  & $-2.53$ \\
558.14 s & 2.4 & 558.04 & $ 0.10$ & 24 & 0.77  & $ 1.51$ \\
579.12   & 0.1 & 579.83 & $-0.71$ & 25 & 1.65  & $-0.02$ \\
603.04   & 0.2 & 602.64 & $ 0.40$ & 26 & 1.03  & $-3.25$ \\
622.00   & 0.3 & 621.89 & $ 0.11$ & 27 & 0.98  & $ 2.42$ \\
643.31 s & 0.5 & 644.05 & $-0.74$ & 28 & 1.77  & $-0.98$ \\
668.09   & 0.3 & 665.84 & $ 2.25$ & 29 & 0.93  & $-2.65$ \\
687.74 s & 0.4 & 686.62 & $ 1.12$ & 30 & 1.43  & $ 2.65$ \\
709.05   & 0.3 & 708.30 & $ 0.75$ & 31 & 1.70  & $-0.76$ \\
729.51 s & 0.3 & 729.32 & $ 0.19$ & 32 & 1.48  & $-2.81$ \\
752.94   & 0.9 & 751.52 & $ 1.42$ & 33 & 1.40  & $ 1.04$ \\
773.74   & 0.3 & 772.65 & $ 1.09$ & 34 & 1.89  & $ 1.92$ \\
791.80   & 6.0 & 794.58 & $-2.78$ & 35 & 2.06  & $-3.98$ \\
814.58   & 0.4 & 815.61 & $-1.03$ & 36 & 1.07  & $-1.37$ \\
838.62 s & 0.6 & 837.38 & $ 1.24$ & 37 & 2.38  & $ 4.41$ \\
861.72   & 0.5 & 860.89 & $ 0.83$ & 38 & 2.20  & $-4.61$ \\
883.67   & 6.0 & 880.45 & $ 3.22$ & 39 & 1.42  & $-2.15$ \\
903.19 s & 0.7 & 902.38 & $ 0.81$ & 40 & 2.31  & $ 2.21$ \\
925.31 s & 0.3 & 925.30 & $ 0.01$ & 41 & 2.31  & $-0.91$ \\
945.01   & 0.3 & 947.13 & $-2.12$ & 42 & 2.21  & $-3.40$ \\
966.98 s & 0.9 & 967.94 & $-0.96$ & 43 & 1.76  & $-0.75$ \\
988.13   & 0.2 & 989.13 & $-1.00$ & 44 & 2.66  &  $3.69$ \\
\hline  
\hline  
\end{tabular}  
\label{best-fit-b}  
\end{table}  

In this section we repeat  the above  analysis but  this time  taking full
advantage of the updated and augmented set of observed periods of \pg\ 
reported by CEA07.  Indeed,  these authors have identified  76 new
pulsation  modes, increasing  to  198 the  total  number of  pulsation
periods of \pg. We reproduce in the second and fourth columns of Table
\ref{periods} the subset of $\ell= 1, m= 0$  and $\ell= 2, m= 0$ periods 
we use here in  our  period fits.  They correspond to Tables 4 
and 12 of CEA07. In our Table \ref{periods}, a letter ``s'' 
means a Confidence  Level of 1 or 2. The observational uncertainties in the 
values of the periods are between $\sim 0.002$ and $\sim 1$ s, 
with an average of $\sim 0.3$ s. \footnote{In the derivation 
of the quoted errors in the observed periods we are assuming that (1) 
each period corresponds to a \emph{real} eigenmode of the star, and (2) the 
indexes $k, \ell, m$ are correctly assigned.}

We note that the pulsation periods of \pg\ are changing with  
rates up to $\sim \pm 30$ ms/yr (Costa \& Kepler 2007 [CK07]). So, 
between 1983 and 2002, the periods could have experienced variations 
up to $0.6$ s. Thus, for modes present in more than one 
periodogram, the periods in Tables 4 and 12 
of CEA07 (and also in our Table \ref{periods}) are the 
\emph{average values}.  
At variance with this, the periods included in Table 6 of CEA07
are the values at the epoch  of the largest amplitude of 
each mode. Thus, these periods correspond to different annual data 
sets. In our period fit procedure below ---and also in the analysis of the 
period spacing of Sect. \ref{mode-trapping}--- we shall employ the set 
of the average values of periods (Tables 4 
and 12 of CEA07) because the use of periods from 
different epochs (Table 6 of CEA07) seems inappropriate.

Regarding the reliability of the  detected modes, CEA07
classified the  possible pulsation  periods by their  confidence level
(CL),  in six  different levels  of decreasing  probability from  1 to
6. The six  probability levels are listed  in Table  2 of  CEA07.

The  results for the $\ell=  1$  period fit are displayed  in the  
left panel of Fig.\ref{cl123456}.   As  previously, a  quite 
prominent peak  in $(\chi^2)^{-1}$ is found, corresponding to the 
$0.565 M_{\odot}$ model
at $T_{\rm eff}=  127\,680$ K. Remarkably, the period  match turns out
be of an excellent quality, with $\overline{\delta \Pi_i}= 1.03$ s and
$\sigma_{_{\delta \Pi_i}}= 1.7$  s, in spite of the  large number (29)
of $\ell= 1$  fitted periods. The recurrent presence  of this solution
throughout all of our period  fits is striking and indicates that this
PG1159  model   could  constitute  a   meaningful  asteroseismological
solution for \pg.

The  results  of  the  period  fit  for $\ell=  2$  (centre  panel  of
Fig. \ref{cl123456}) are  not very different from those  of the $\ell=
2$  period fits previously  discussed.  However,  in the  present case
there exists  a very pronounced peak  in $(\chi^2)^{-1}$ corresponding
to a $0.565 M_{\odot}$ model at $\approx 167\, 000$ K, associated with
a period match characterized by $\overline{\delta \Pi_i}= 0.60 $ s and
$\sigma_{_{\delta \Pi_i}}= 0.33$ s. Unfortunately,  since this  period 
fit relies on the observed $\ell= 2$  periods only, which generally  
have more uncertain identifications of $\ell$ and $m$ than  dipole
periods\footnote{In particular, the $m$  identification for $\ell= 2$ 
is insecure,  because generally  not all  the components  of the
quintuplets are detected.  This leads to large  uncertainties in the
precise value of $m$ for a given mode.}, and because this solution 
does not also appear for the dipole modes ---for which we are confident 
of the $\ell$ and $m$ values--- we must discard this model as
a  meaningful asteroseismological  solution for  \pg.  Other important
peak seen in $(\chi^2)^{-1}$ is  located at $T_{\rm eff} \approx 100\,
000$ K for $M_*= 0.609  M_{\odot}$. In decreasing order of importance,
we found other peaks at $T_{\rm eff} \approx 138\, 000$ K ($M_*= 0.565
M_{\odot}$),  $T_{\rm   eff}  \approx   112\,  500$  K   ($M_*=  0.589
M_{\odot}$),   $T_{\rm  eff}   \approx  83\,   500$  K   ($M_*=  0.664
M_{\odot}$),  and  $T_{\rm eff}  \approx  271\,  000$  K ($M_*=  0.741
M_{\odot}$).

All of  the significant peaks described  above are ironed  out when we
consider  a  period  fit for  the  $\ell=  1$  and $\ell=  2$  periods
simultaneously,  as   can  be   seen  in  the   right  panel   of  the
Fig.  \ref{cl123456}.   Again, no  clear  solution emerges  satisfying
simultaneously  both the  $\ell= 1$  and  $\ell= 2$  sets of  observed
periods, although there is a wide bump just below $\sim 140\,000$ K 
for the sequence with $M_*= 0.565 M_{\odot}$.

Next, we  repeated the period fits  described above,
but this time by considering only  the most probable modes of the data
set, i.e.   the modes with CL=  1 or 2.   Modes with CL= 1  have large
amplitudes  and appear  in one  or more  of the  periodograms  of CEA07, 
whereas modes with  CL= 2 have lower  amplitudes but even  above the detection
limit, and  appear in  two or more  periodograms.  The periods  of the
modes with CL= 1 or 2  according CEA07 are labelled with
a letter  ``s'' in  Table \ref{periods}.  Note  that in this  case the
periods to be fitted are drastically  reduced in number, from 29 to 14
for $\ell=  1$ and  from 26  to 7 for  $\ell= 2$.   

\begin{figure*}  
\centering  
\includegraphics[clip,width=400pt]{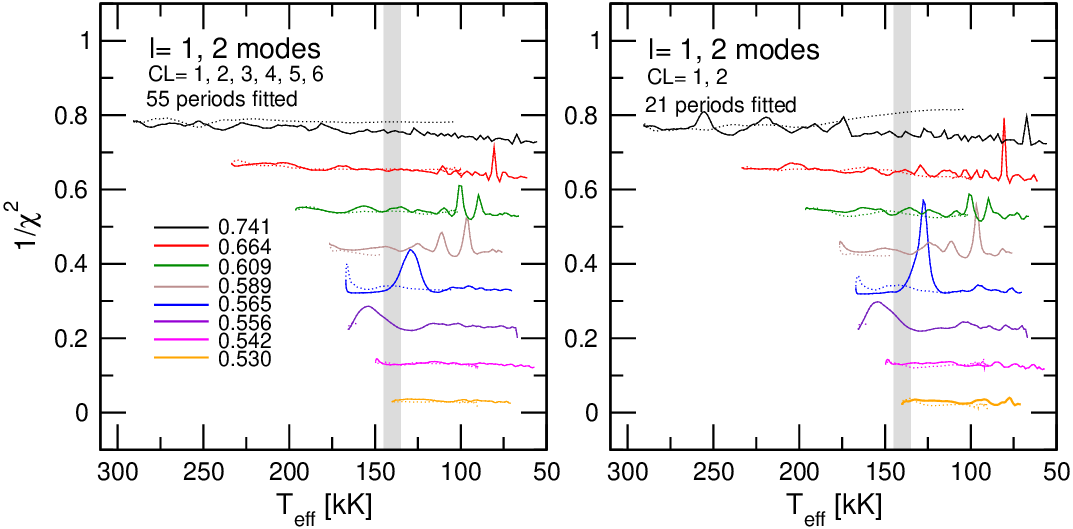}  
\caption{The inverse of the quality function corresponding to 
period fits  considering $\ell=  1,  2$ modes  simultaneously.
Observed $\ell= 1$ ($\ell= 2$) periods are compared with
theoretical $\ell= 1$ ($\ell= 2$) periods. The left panel 
corresponds to a period fit that includes all  
of the  observed $m=  0$ periods (having a Confidence Level of 1, 2, 3, 4, 5, 
or 6) extracted from Tables 4  and  12 of CEA07. The right panel
corresponds to a period fit in which only the high confidence periods 
(those having CL= 1 or 2) are employed. [Color
figure only available in the electronic version of the article]}
\label{fixed}  
\end{figure*}  

The results of our analysis are presented in Fig. \ref{cl12}.  
Notably, for the $\ell= 1$ period  fit  we   again  recover  the  persisting  
solution  with  $M_*= 0.565  M_{\odot}$  and  $T_{\rm eff}=  127\,680$
K. Remarkably, the quality of this fit for
\pg\ is much better than of those previously discussed in this paper 
for  $\ell=  1$ modes,  with  $\overline{\delta  \Pi_i}=  0.64$ s  and
$\sigma_{_{\delta  \Pi_i}}=  0.75$ s.  A  comprehensive comparison of  the
observed periods with  the theoretical periods for this  fit is
provided in Table \ref{best-fit-b}.

\begin{figure*}  
\centering  
\includegraphics[clip,width=500pt]{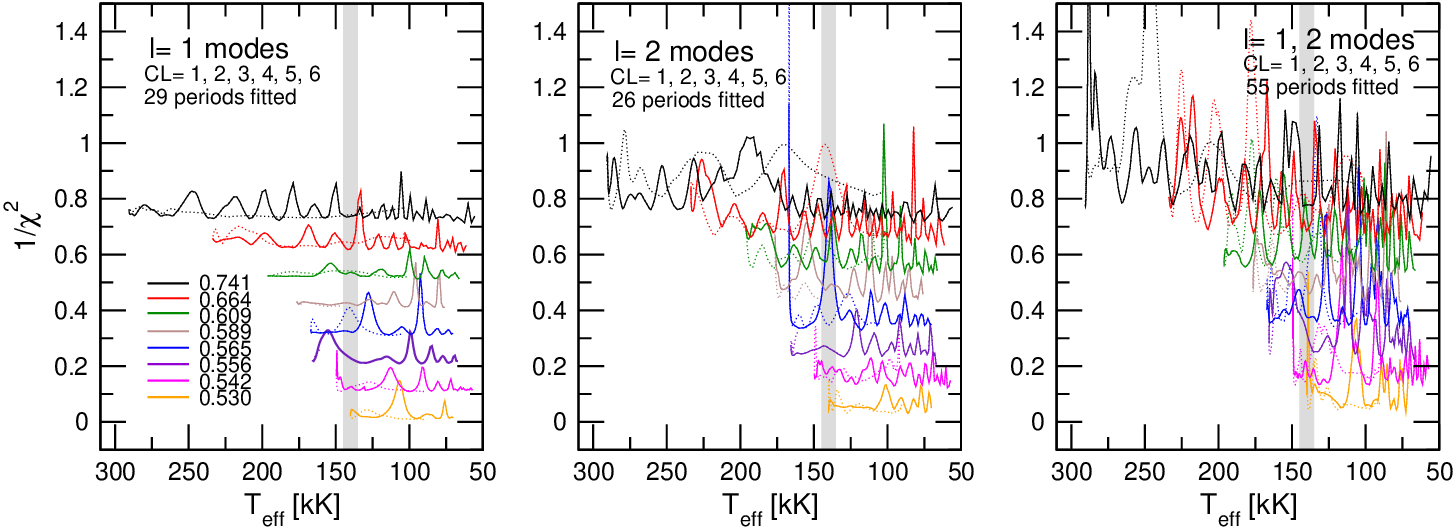}  
\caption{Same as Fig. \ref{cl123456}, but for the case in which 
the observed $m= 0$  periods are weighted with the square of the 
corresponding amplitude 
in the period fit procedure. [Color
figure only available in the electronic version of the article]}
\label{weighted}  
\end{figure*}  

For the $\ell= 2$ period fit  we obtain similar results than above
(see centre panel of Fig. \ref{cl12}). In
particular,  the acute  peak  corresponding to  the $0.565  M_{\odot}$
model at $\approx 167\, 000$ K persists, and in the present case it is
characterized    by   $\overline{\delta    \Pi_i}=    0.71$   s    and
$\sigma_{_{\delta \Pi_i}}= 0.75$ s. The new feature is the presence of
a  pronounced  peak  in  $(\chi^2)^{-1}$  corresponding  to  a  $0.664
M_{\odot}$ model at $T_{\rm eff} \sim 147\, 000$ K, before the turning
point in  $T_{\rm eff}$ is reached.  The peak at  $T_{\rm eff} \approx
100\,  000$ K  for  $M_*= 0.609  M_{\odot}$,  on the  other hand,  has
suffered from a notable reduction of its amplitude.

Again we are unable  to found a unambiguous solution for the
case in which the $\ell= 1$ and $\ell= 2$ sets of observed periods are
compared simultaneously  with $\ell= 1, 2$  theoretical periods.  This
is evident from the right panel of Fig. \ref{cl12}.

\subsubsection{Fixing the $\ell$ value of the observed periods}
\label{fixing}

Here we perform additional period fits 
for $\ell= 1$ and $\ell= 2$ simultaneously but forcing the $\ell= 1$ 
observed modes to be fitted by $\ell= 1$ theoretical modes, and 
the $\ell= 2$ observed modes to be fitted by $\ell= 2$ theoretical 
modes. In other words, we assume that the $\ell$ 
values adopted by CEA07 for the observed periods are correct. 
Fig. \ref{fixed} displays in the left panel the results for a 
period fit considering all the periods (CL= 1, 2, 3, 4, 5, or 6) and 
in the right panel the results for a period fit in which only 
the CL= 1 or 2 are taken into account. Note that in both cases we found 
again the best-fit model with $M_*= 0.565 M_{\odot}$ and 
$T_{\rm eff}= 127\,680$ K. The fact that both period fits ---one 
involving 55 periods and the other only 21--- lead to similar results
lends confidence to this asteroseismological solution. 

\subsubsection{Weighting the period fits}
\label{weighting}

We have finally performed  period fits  by using  the observed  periods of
CEA07 with CL=  1, 2, 3, 4, 5  and 6, but weighting each
period  in the  quality function  $\chi^2$ with the corresponding amplitude. 
Specifically,  the quality function now reads:

\begin{equation}
\chi^2(M_*, T_{\rm
eff})= \frac{\sum_{i=1}^n \min[(\Pi_i^{\rm  O}- \Pi_k)^2] A_i^2} 
{\sum_{i= 1}^n A_i^2}, 
\label{chi2-weighted}
\end{equation}

\noindent where we have adopted the  amplitudes $A_i$ of the observed modes
---extracted from Tables 4 and 12 of CEA07--- as the weights.  
In this  way, the period fits
are more influenced by modes with large amplitudes than by modes with
low amplitude.   In other words,  modes of large amplitude  must match
better the theoretical model than  modes of small amplitude. Note that
if we set $A_i= 1$ for  $i= 1,\cdots ,n$, i.e., assume the same weight
for all of the modes, Eq. (\ref{chi2-weighted}) reduces  
to Eq. (\ref{chi2}), which we used in the period fits of the previous 
sections. 

As before, in the period fits  the harmonic degree was
allowed to  adopt the values $\ell=  1$ or $\ell= 2$.  The results are
shown in Fig. \ref{weighted}. Unfortunately, no apparent  
solution is found, neither in the case of $\ell= 1$ (left panel) 
or $\ell= 2$ (centre panel) period fits, nor in the case 
of period fits to $\ell= 1$ and $\ell= 2$ modes simultaneously  (right panel). 
This is somehow expected since by using Eq. (\ref{chi2-weighted}) 
the period fit is dominated by only a few large-amplitude modes 
(those with periods 390.30, 452.44, 
517.16, 558.14, 791.80, and 883.67 s), whereas the remaining modes, which 
have notably smaller amplitudes, barely  contribute to the weighting 
process. As a result, multiple and almost equivalent minima in 
the $\chi^2$ function are 
obtained, rendering it virtually impossible to find an 
unambiguous asteroseismological solution.       

\begin{table*}  
\centering  
\caption{The main characteristics of \pg.}   
\begin{tabular}{l|cccr}  
\hline  
\hline  
Quantity & Spectroscopy &   Asteroseismology   &  Pulsations    & Asteroseismology        \\  
         &              &   (KB94)             &  (CEA07, CK07) & (this work)            \\  
\hline  
$T_{\rm eff}$ [kK]           & $140 \pm 5^{\rm (a)}$          &  $136$           & ---              & $128^{+8.6}_{-2.6}$ \\  
$M_*$ [$M_{\odot}$]          & $0.54\pm 0.1^{\rm (b)}$        &  $0.59\pm 0.01$  & $0.59 \pm 0.02$  & $0.565^{+0.025}_{-0.009}$ \\   
$\log g$ [cm/s$^2$]          & $7.0 \pm 0.5^{\rm (a)} $       &  $7.38\pm01$     & ---              & $7.42^{+0.21}_{-0.12}$   \\   
$\log (L_*/L_{\odot})$       & $2.73_{-0.55}^{0.41\rm (c)}$   &  $2.29\pm0.05$   & ---              & $2.15 \pm 0.08$   \\    
$\log(R_*/R_{\odot})$        & ---                            &  $-1.60\pm 0.09$ & ---              & $-1.62^{+0.06}_{-0.09}$  \\    
$M_{\rm env}$ [$M_{\odot}$]  & ---                            &  $0.002$         & ---              & $0.017$  \\ 
$r_{\rm c}/R_*$              & ---                            &  $0.60-0.65$     & $0.83 \pm 0.05$  & $0.55$   \\ 
C/He, O/He$^{(*)}$           &  $1.5, 0.5^{\rm (c)} $         & ---              & ---              & $0.81, 0.56$ \\     
\hline
$M_{\rm V}$ [mag]            & ---                            & ---              & ---              & $6.975 \pm 0.4$       \\
$M_{\rm bol}$ [mag]          & ---                            & ---              & ---              & $-0.625 \pm 0.2$       \\
$A_{\rm  V}$ [mag]           & ---                            & ---              & ---              & $0.06^{+0.003}_{-0.004}$ \\ 
$d$  [pc]                    & $800^{+600 {\rm (a)}}_{-400}$  & $440 \pm 40$     & ---     & $363_{-61}^{+73}$  \\ 
$\pi$ [mas]                  & $1.25_{-0.54}^{+1.25}$         & $2.27_{-0.18}^{+0.23}$     & ---              & $2.75_{-0.46}^{+0.56}$ \\ 
\hline
$\dot{R}_*/R_*$ [$10^{-11}$s$^{-1}$]   & ---                  & ---              &  $-0.89\pm 0.2$  &  $-0.009$ \\  
$\dot{T}/T$     [$10^{-11}$s$^{-1}$]   & ---                  & ---              &  $-1.84\pm 0.04$  &  $0.0004$ \\  
\hline
$k$ for $\Pi= 517$ s & --- & 22 & 20 & 22 \\
\hline  
\hline  
\end{tabular}  
\label{model}  
  
{\footnotesize  Note: $(*)$Abundances by mass}  
  
{\footnotesize  References: (a)  Werner et al. (1991);   
(b) Miller Bertolami \& Althaus (2006); (c) Werner \& Herwig (2006).}  
\end{table*}

\subsection{The asteroseismological best-fit model}  
\label{best-fit}  
 
The results  described in the  last sections strongly  suggest the
existence of a significant asteroseismological solution
for \pg\ corresponding  to a model with a stellar  mass of $M_*= 0.565
M_{\odot}$  and  $T_{\rm  eff}=   127\,680$  K.   We  arrive  at  this
conclusion by considering mostly  $\ell= 1$ period fits, although this
model also  provides a  good period  fit for $\ell=  1$ and  $\ell= 2$
modes  simultaneously  if  we  force observed $\ell= 1$ periods to be 
compared with theoretical $\ell= 1$ periods, and observed 
$\ell= 2$ periods to be compared with theoretical $\ell= 2$ periods
(Sect. \ref{fixing}). Also, we found a good period fit for this model 
when we consider  the  observed periods  with  secure identification     
of     WEA91, allowing the periods to be  
$\ell= 1$ or $\ell= 2$ without constraining any to be one or the other 
(Table \ref{best-fit-a}). This  model will  
be  adopted as  the ``best-fit model'' representative of  
\pg \footnote{It is important to
stress that period  fits using only $\ell= 2$ modes show interesting possible
solutions, in particular the $0.565 M_{\odot}$ model at $\approx 167\,
000$ K described in the  last section.  However, we must discard these
solutions because  they exclude the  $\ell= 1$ modes (which  have more
secure identifications than the $\ell= 2$ modes).}. 

The quality of our $\ell= 1$ period fits for \pg\ using the periods of
CEA07 is characterized by  $\overline{\delta \Pi_i}= 1.02$ s (when we
consider modes with CL= 1, 2,  3, 4, 5, or 6; see Fig. \ref{cl123456})
and $\overline{\delta \Pi_i}=  0.62$ s (if we take  into account modes
with CL= 1  or 2 only; see Fig. \ref{cl12}). These  period fits are of
better quality  than those obtained  by KB94 and C\'orsico  \& Althaus
(2006) ($\overline{\delta
\Pi_i}= 1.19$ s and $\overline{\delta \Pi_i}= 1.79$ s, respectively)
for PG  1159-035. The $\ell= 2$  periods of our best-fit  model 
do not match the  observed  quadrupole  period spectrum as accurately  
as the dipole periods do. In spite of this fact, 
the  best-fit model  yields an  average of  the computed
period spacings for $\ell= 2$ of $\overline{\Delta \Pi_{\ell= 2}}= 12.
61$ s, in good agreement with the observed value of $\Delta
\Pi_{\ell=  2}^{\rm O}= 12.38 \pm  0.01$ s.   This property renders
consistency to the derived best-fit model.

As in our previous asteroseismological 
analysis of GW Vir stars, we  are able to get  a PG1159 model that  
reproduces  the period
spectrum observed in the star  under study without tuning ad hoc the
value  of structural  parameters such  as the  thickness of  the outer
envelope, the  surface chemical  abundances or the  shape of  the core
chemical  profile  which,  instead,  are  kept  fixed  at  the  values
predicted by the evolutionary computations.

Table  \ref{best-fit-b} shows  a  detailed comparison  between the  29
$\ell= 1$ CEA07 observed periods  ($\Pi_i^{\rm O}$) and
the  theoretical  $\ell=  1$   periods  ($  \Pi_k$)  of  our  best-fit
model. Note that if we consider all  the 29 periods (CL= 1, 2, 3, 4, 5
or 6),  we have $|\delta \Pi_i|\leq 1$  s for 17 modes  ($59 \%$), $1\
\mbox{s}<|\delta  \Pi_i|  \leq  2$  s  for  7  modes  ($24  \%$),  $2\
\mbox{s}<|\delta \Pi_i| \leq 3$ s  for 4 modes ($14 \%$), and $|\delta
\Pi_i| >  3$ s  for 1  mode ($3.5 \%$).  However, if  we adopt  the 14
periods with  the highest level of  probability only (CL= 1  or 2), we
obtain $|\delta \Pi_i|\leq 1$ s for 11 modes ($\approx 80 \%$) and $1\
\mbox{s}<|\delta \Pi_i| \leq 2$ s for 3 modes ($\approx 20 \%$).

The fifth column of Table \ref{best-fit-b} presents the 
radial order $k$ for the modes of our best-fit model. The $k$ values are 
in complete agreement with those of the best-fit model of KB94, and 
also in agreement (to within $\Delta k= \pm 2$) with the values  
proposed by CEA07. As a reference, we show the radial order of the 
mode with the largest amplitude ($\Pi \sim 517$ s), according 
to the different authors, in the last row of Table \ref{model}. 

In Table \ref{model} we show the main features of our best-fit model 
and also the parameters of \pg\ extracted from other published studies. 
Specifically, the second column    
corresponds to spectroscopic results, the third column presents results from 
the asteroseismological analysis of KB94, the fourth column corresponds 
to the pulsation results of CEA07 and Costa \& Kepler (2007) (CK07), and the 
fifth column shows the characteristics of the asteroseismological model 
of this work. The quoted uncertainties in the stellar mass 
and the effective temperature of our best fit model 
($\sigma_{\rm M_*}$ and $\sigma_{T_{\rm eff}}$) resulting 
from the period fit procedure are calculated according to 
the following expression, derived by Zhang et al. (1986):

\begin{equation}
\sigma_i^2= \frac{d_i^2}{S-S_0}
\end{equation}

\noindent where $S_0= \chi^2(M_*^0, T_{\rm eff}^0)$ 
is the absolute minimum of $\chi^2$ which is reached at 
($M_*^0, T_{\rm eff}^0$) corresponding to the best-fit model, and 
$S$ the value of $\chi^2$ when we change the parameter $i$ 
(in our case, $M_*$ or $T_{\rm eff}$) by an amount $d_i$ 
keeping fixed the other 
parameter. The quantity $d_i$ can be evaluated as  the minimum  
step in the grid of the parameter $i$. Specifically, we use the $\ell= 1$ 
period fit shown in the left panel of Fig. \ref{cl12}, 
that is, by considering the 14 dipole periods with
CL= 1 or 2 measured by CEA07. 
We have $d_{T_{\rm eff}} \equiv \Delta T_{\rm eff}\sim  1000$ K
and $d_{M_*} \equiv \Delta M_*$ in the range $0.009-0.024 M_{\odot}$. 

Our  best-fit model  has a  stellar  mass of  $M_*= 0.565  M_{\odot}$,
which is $\sim 1.9 \%$ smaller than  the   value  derived from  
the  average of  the computed period spacing, $M_*  \sim 0.576 M_{\odot}$, 
and $\sim 3 \%$ lower than that inferred  from the 
asymptotic period spacing, $M_*\sim 0.581 M_{\odot}$ 
(see  \S \ref{period-spacing}).   On the  other hand,
CEA07 have inferred  a value of the stellar mass of  \pg\ by using an
interpolation  formula to  the period  spacing derived  by  Kawaler \&
Bradley  (1994) on  the basis  of a  large grid  of  artificial PG1159
models  in  the  luminosity  range $1.6  \lesssim  \log(L_*/L_{\odot})
\lesssim 3.0$.  These authors obtain  a rather high value of $0.59 \pm
0.02  M_{\odot}$,  in line  with  the early  determinations
of WEA91 and KB94 ($0.586 \pm 0.003 M_{\odot}$ and 
$0.59 \pm 0.01 M_{\odot}$, respectively), but in disagreement 
--- about $4.5 \%$ larger --- with the mass of our best-fit model.

\begin{figure}
\centering
\includegraphics[width=220pt]{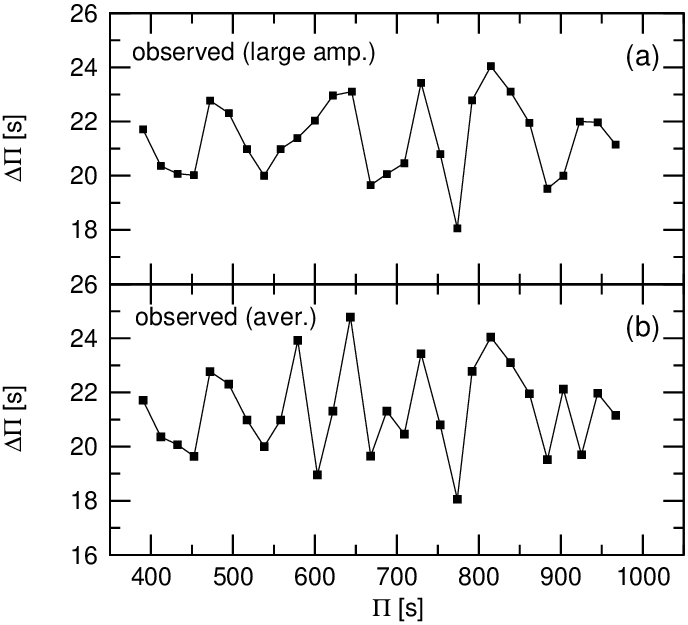}
\caption{Distribution of the forward period spacing 
$\Delta \Pi$ for the case of the observed periods according to 
Table  6 ---panel (a)--- and according to Table 4 ---panel (b)--- of CEA07.}  
\label{delp-cea07}
\end{figure}

The $M_*$ value of our  best-fit model is somewhat
higher  than the  spectroscopic mass  of $0.54  M_{\odot}$  derived by
Miller Bertolami \&  Althaus (2006) (see also Dreizler  \& Heber 1998)
for \pg.  Note that  a discrepancy between the asteroseismological and
the  spectroscopic  values of  $M_*$  is  generally encountered  among
PG1159 pulsators  (see C\'orsico et  al. 2006, 2007ac and Althaus et al. 
2007).  Until  now, the
asteroseismological  mass of  \pg\ has  been about  $9 \%$ larger
($\Delta  M_*  \approx 0.05 M_{\odot}$)  than the  spectroscopic
mass.   In light of  the best-fit  model derived  in this  paper, this
discrepancy is  reduced to  less than about $5 \%$ ($\Delta M_*
\approx 0.025 M_{\odot}$).

The location of the best-fit model in the 
$\log T_{\rm eff} - \log g$ diagram is displayed in Fig. \ref{teff-g}. 
Note that the model is characterized by a $T_{\rm eff}$ 
somewhat lower than the spectroscopic value. Also,   
the model lies certainly slightly outside the $\ell= 1$ GW Vir 
instability domain, at odds with the observational evidence. 

\subsection{Period spacing and mode trapping}
\label{mode-trapping}

The mode-trapping properties of PG1159 stars
have been discussed at length by KB94 and C\'orsico \& Althaus (2006);
we refer the  reader to those works for details.  We will consider the
observed $\ell=  1$ forward period spacing  ($\Delta \Pi= \Pi_{k+1}-
\Pi_k$)  taking advantage  of the  uninterrupted string  of  29 dipole
periods  with consecutive  radial  orders  from $k=  14$  to $k=  42$,
according CEA07.

In panel (a)  of Fig.  \ref{delp-cea07} we show  the $\Delta \Pi$ vs
$\Pi$  diagram for  the observed  periods  according to  Table 6  of
CEA07. The uncertainties in the observed period spacings 
are between $0.01$ and $1.2$ s, with an average value of $0.4$ s.
As we  mentioned at the beginning of Sect. \ref{costa}, 
the periods of \pg\  are changing with time. So, the values included 
in Table  6 of CEA07 are the periods corresponding
to the epoch  of the largest amplitude of each mode.   This diagram is used
by  those  authors to  identify  five  modes  trapped in  the  He-rich
envelope of \pg.  They arrive  at this conclusion under the assumption
that trapped  modes appear as points  of minimum in  this diagram. 
By  assuming  the presence  of a  \emph{single}
chemical interface,  CEA07 derive $r_{\rm c}/R_* \approx  0.83$, 
where $r_{\rm c}$
is the  radial coordinate  of the O/C/He  chemical interface.   In our
best-fit  model this  chemical  interface is  located  at markedly  deeper
layers ($r_{\rm c}/R_* \approx 0.55$), in  agreement with the models of KB94
($r_{\rm c}/R_* \approx 0.60-0.65$).

\begin{figure}
\centering
\includegraphics[width=230pt]{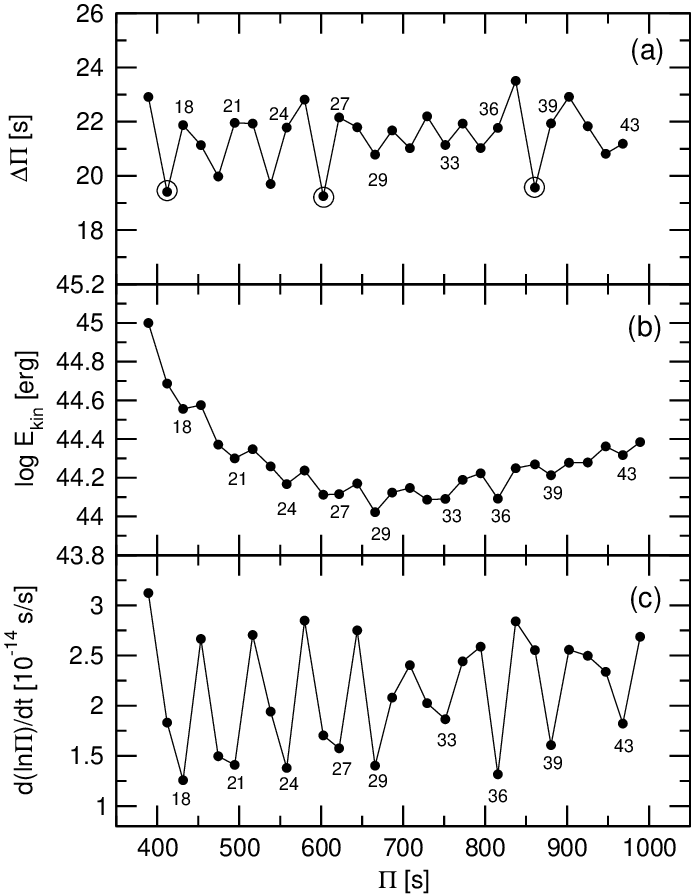}
\caption{Panel (a): distribution of $\Delta \Pi$ for the best-fit
model, the primary minima being marked with circles.  
Panel (b): the  computed distribution of the kinetic energy.  
Panel (c):  the theoretical values  of the relative  rates of 
period  change. Modes trapped in the outer envelope model are 
labelled with their radial order $k$.}  
\label{paneles}
\end{figure}

Note that  some periods used to construct the  diagram of panel (a)
in Fig.  \ref{delp-cea07} (extracted from Table 6 of CEA07) 
slightly differ from the periods included in
Table 4 of CEA07 (and in our Tables \ref{periods} and
\ref{best-fit-b})  because, as stated before, they are the average values
of periods on several annual data sets. The slight differences in some 
of the 
periods are translated into  appreciable differences in some values of
the period spacing, as can be seen in panel (b) of Fig.
\ref{delp-cea07}, where we plot the same diagram but on the basis of
the periods given in Table 4 of CEA07. Since the period fits performed
in the present  work are based on these  periods, we shall concentrate
on the diagram  of panel (b) of  Fig. \ref{delp-cea07}. 
It  can be compared with
panel  (a) of  Fig.  \ref{paneles},  in  which the  $\Delta \Pi$  vs
$\Pi$ diagram  for the  best-fit model is  
displayed\footnote{The uncertainties in $T_{\rm eff}$ and  $M_*$ of the 
best-fit model (see Table \ref{model}) lead to uncertainties in 
the computed $\Delta \Pi$, with an average value of 
$\overline{\sigma_{{}_{\Delta \Pi}}} \approx 1$ s. 
In the quoted errors of  $T_{\rm eff}$ and $M_*$ of the 
best-fit model we are neglecting possible uncertainties 
in the stellar evolution and pulsation calculations. Thus, the 
quoted error in $\Delta \Pi$ must be taken as lower limit.}.
Note  that, in
spite  of the  excellent match  between the  observed periods  and the
theoretical  periods of the  best-fit model  (Table \ref{best-fit-b}),
the pertinent  period spacing distribution looks  somewhat different.
The disagreement  has its  origin in  the fact  that small  differences in
period ---because  the period match  is not perfect---  are translated
into large differences in  period spacing. Besides the disagreement
between individual values of $\Delta  \Pi$, we note
that  generally the amplitude  of  the observed  period  spacing are  
remarkably larger than that of the best-fit model. This suggests that
the  chemical interfaces present  in \pg\  could be  considerably 
steeper than predicted by our PG1159 modeling.

Notwithstanding the  mentioned differences between  the period spacing
distributions in \pg\  and in our best-fit model,  they share a common
feature: both exhibit primary  and secondary minima, characteristic of
mode trapping due to the presence of
\emph{more than one} chemical interfaces. In fact, our best-fit 
model has two chemical transition regions: the inner interface of O/C
and  the  more  external  interface  of O/C/He.  As  we  have  already
demonstrated (see,  e.g., C\'orsico \&  Althaus 2006 and  C\'orsico et
el. 2007ac),  for periods below $\approx 650-700$ s the mode-trapping 
features  of our PG1159 models are induced mostly by the chemical 
gradient at the O/C/He interface, the O/C interface being 
more relevant for longer periods.   We conclude  that  the presence  of  
primary and  secondary minima in  the $\Delta \Pi_k$ pattern of  \pg\ is an 
indication that the star could be characterized by more than one 
chemical interfaces, possibly two\footnote{This conclusion is valid 
even if we consider the observational uncertainties in $\Delta \Pi$.}.

Regarding the identification of trapped  modes by means of the $\Delta
\Pi$ vs $\Pi$ diagram  ---which constitutes the main observational
tool  of mode  trapping  in  PG1159 stars---  we  emphasize  that
generally a  trapped mode with radial  order $k$ is  associated with a
minimum $\Delta \Pi$ which corresponds  to a radial order $k \pm 1$,
and in some cases to $k$. This is illustrated in panel (b) of 
Fig. \ref{paneles},
in  which  we  show  the   kinetic  energy  distribution  $E_{\rm
kin}$\footnote{The usual normalization
of  the radial  eigenfunction  ($y_1= \xi_r/r=  1$  at the  stellar
surface) is assumed.},  
of  our best-fit model.  Local minima in the  $E_{\rm kin}$
distribution  correspond   to  trapped  modes,   because  these  modes
propagate  mainly in  the low  density  regions of  the outer  He-rich
envelope. Note that, for instance,  the minimum of $\Delta \Pi$ with
radial order  $k= 23$ is  related to  a minimum in  $E_{\rm kin}$  of the
trapped mode  with $k=  24$.  Modes trapped  in the envelope  are also
characterized  by minima  in $\dot{\Pi}/\Pi$,  because they  should be
more strongly  sensitive to the  effects of the  surface 
contraction ---that induces a  secular decrease of the periods--- than
untrapped modes ---which are more  affected by cooling. This effect is
well illustrated in panel (c) of Fig. \ref{paneles}.

\subsection{Period changes and the cooling rate of \pg}
\label{period-changes}

In Table  \ref{best-fit-b} we show the  rate of period  change for the
modes of our best-fit  model (sixth column).  Our calculations predict
all of the pulsation periods  to {\it increase} with time ($\dot{\Pi}>
0$),  in  accordance  with  the decrease  of  the  Brunt-V\"ais\"al\"a
frequency in the core of  the model induced by cooling.  This property
is shared  by all of  the PG1159 models  located below the  lower thin
dotted  line  in  Fig.  \ref{teff-g}.   Note  that  at  the  effective
temperature of  our best-fit model,  cooling possibly has  the largest
effect on  $\dot{\Pi}$, but gravitational  contraction ---which should
result in a  {\it decrease} of periods with  time--- is not negligible
and could affect the $\dot{\Pi}$ values, in particular for the case of
modes {\it trapped} in the envelope (see Fig. \ref{paneles}).

Our  best fit model does  not reproduce  the  recent measurements  of CK07, 
which indicate  that the pulsation modes of \pg\
have \emph{positive  and negative}  values of $\dot{\Pi}$.   
Our PG1159  models having a  mix of positive and  negative $\dot{\Pi}$
values   are    located   between    the two thin   dotted    lines   in
Fig. \ref{teff-g}. Remarkably, the measurements of CK07 are consistent
with the spectroscopic location of \pg\ on the $\log T_{\rm eff}- \log
g$  diagram. Thus,  to be  consistent with  the findings  of  CK07 our
best-fit  model for  \pg\ 
should be located  at  earlier evolutionary stages.

CK07 report  values of  the rate  of period change  in \pg\  that are
generally more than one order of magnitude larger than the theoretical
$\dot{\Pi}$ values  characterizing our best-fit  model. In particular,
they report $\dot{\Pi}= (+18.2 \pm 0.8)
\times  10^{-11}$ s/s for the large amplitude $m= 0,\ \ell= 1$ mode 
with period  517.1 s, excessively  larger than the value  predicted by
our  computations for  that mode  ($\dot{\Pi}= 1.40  \times 10^{-11}$
s/s).  In  addition, CK07  have been able  to measure the  second order
rates of period change ($\ddot{\Pi}$)  for eight modes in \pg.  Again,
the  measured  values  ($|\ddot{\Pi}|  \sim  0.01-9  \times  10^{-19}$
ss$^{-2}$)  are  strikingly larger  than  our theoretical  predictions
($|\ddot{\Pi}| \sim 0.1-5 \times 10^{-23}$ ss$^{-2}$).

Finally, by using the $\dot{\Pi}/\Pi$ values measured for the ($m= 0$)
517.1  s and  the ($m=  +1$) 516.0  s modes,  CK07 
roughly estimate for  \pg\ a
contraction    rate   of    $\dot{R}_*/R_*=    -8.9\pm   0.2    \times
10^{-12}$s$^{-1}$ and a cooling rate of $\dot{T}/T= -1.84 \pm 0.04 \times
10^{-11}$s$^{-1}$, orders of magnitude larger than  our theoretical  
predictions for
the best-fit  model, of $\dot{R}_*/R_*=  -9.0 \times 10^{-14}$s$^{-1}$
and $\dot{T}/T= 4.0 \times 10^{-15}$s$^{-1}$.

\subsection{The asteroseismological distance and parallax of \pg}
\label{distance}

We employ  the luminosity of our  best-fit model to  infer the seismic
distance of  \pg\ from  the Earth. First,  we compute  the bolometric
magnitude  from the  luminosity  of  the best-fit  model  by means  of
$M_{\rm  bol}= M_{\odot  {\rm bol}}  - 2.5  \log(L_*/L_{\odot})$, with
$M_{\odot  {\rm bol}}= 4.75$  (Allen 1973).  Next, we  transform the
bolometric magnitude  into the absolute magnitude,  $M_{\rm v}= M_{\rm
bol} - {\rm  BC}$, where ${\rm BC}$ is  the bolometric correction. For
\pg\ we  adopted ${\rm BC}= -7.6  \pm 0.2$ from Werner  et al. (1991).
We  account for the  interstellar absorption,  $A_{\rm V}$,  using the
interstellar extinction  model developed by  Chen et al.   (1998).  We
compute the seismic distance $d$ according to the well-known relation:
$\log d = \frac{1}{5} \left[ m_{\rm v} - M_{\rm v} +5 - A_{\rm V}(d)
\right]$ where  the apparent  magnitude is $m_{\rm v}=  14.84$. 
The interstellar  absorption $A_{\rm  V}(d)$ varies non  linearly with
the  distance and  also depends  on  the Galactic  latitude ($b$)  and
longitude  ($\ell$). For  the  equatorial coordinates  of \pg\  (Epoch
B2000.00,  $\alpha= 12^{\rm  h}\  1^{\rm m}\  46^{\rm s}.1$,  $\delta=
-3^{\circ}\ 45'\ 39''$) the corresponding Galactic coordinates are $b=
-56^{\circ}\ 51'\  55''.44$ and  $\ell= 279^{\circ}\ 49'\  10''.2$. We
solve for  $d$ and  $A_{\rm V}$ iteratively  and obtain a  distance $d
\sim 360$  pc and  an interstellar extinction  $A_{\rm V}  \sim 0.06$.
Note  that  our  distance  is  a factor of more than two smaller    
than  the  estimation  of $800^{+600}_{-400}$  of Werner  et al.  (1991) 
and  with  its accuracy substantially improved.  Finally,  
our calculations predict a parallax of $\pi \sim 2.8$ mas.

\section{Discussion and conclusions}  
\label{conclusions}  

In this paper we carried out a comprehensive asteroseismological study of
\pg, a $g$-mode pulsator that defines the
class of pulsating PG1159 stars --- the GW Vir variables--- and the PG
1159  spectral class  of  hydrogen-deficient stars.   Our analysis  is
based  on  the full  PG1159  evolutionary  models  presented in 
Althaus  et  al.
(2005), Miller  Bertolami \& Althaus (2006), C\'orsico  et al.  (2006)
and C\'orsico et al. (2007c).  These models represent a solid basis to
analyze the evolutionary  and pulsational status of GW  Vir stars like
\pg.  

We first took advantage of the strong dependence of the period spacing
of variable PG1159  stars on the stellar mass, and  derived a value of
the  mass in the  range $0.577-0.585  M_{\odot}$ by  comparing 
the observed mean period spacing $\Delta
\Pi^{\rm O}$  with the  asymptotic period spacing  of our  models.  We
also compared $\Delta
\Pi^{\rm  O}$ with  the  computed period  spacing  averaged over  the
period  range  observed in  \pg,  and derived  a  value  in the  range
$0.561-587 M_{\odot}$.  Note that in  both derivations of  the stellar
mass we  made use of  the spectroscopic constraint that  the effective
temperature of the star should be $\sim 140$ kK.

Next, we adopted a less  conservative approach in which the individual
observed  pulsation  periods   alone  ---i.e.,  ignoring  ``external''
constraints such  as the spectroscopic  values of the  surface gravity
and     effective    temperature---     naturally    lead     to    an
``asteroseismological''   PG1159   model  that   is   assumed  to   be
representative of  the target star. Specifically,  the method consists
in looking  for the model  that best reproduces the  observed periods.
The period fits were made on a grid of PG1159 models with a quite fine
resolution in effective temperature although admittedly  coarse in 
stellar  mass.  We  employed   the  period  data  from  WEA91 
and also the more modern --- and enlarged--- period data of
CEA07.  We  considered period fits taking  into account
observed  $\ell= 1$  modes only,  observed $\ell=  2$ modes  only, and
observed  $\ell= 1$  and $\ell=  2$ modes  simultaneously. We  found a
persisting  and  significant  asteroseismological  solution  for  \pg\
corresponding to a model with a stellar mass of $M_*= 0.565 M_{\odot}$
and $T_{\rm  eff}= 127\,680$ K whose $\ell= 1$ periods match 
the observed periods identified as $\ell= 1$ with an  
average  of  the  period   differences  
(observed  versus theoretical) of  only $0.64$ s (for  14 $\ell= 1$  
periods fitted) and
$1.03$  s (for  29  $\ell=  1$ periods  fitted). 
Also, for this ``best-fit model'' we obtain a good period match when we 
consider a period fit for dipole and quadrupole modes 
\emph{simultaneously} if we constrain the $\ell$ value of the input 
periods from the outset. 

The main  results of  this work are  summarized in  Fig. \ref{teff-g}
and Table \ref{model},
where the best-fit  model solution as well as  the inferences from the
period spacing analysis are  displayed together with the spectroscopic
inferences.   Note  that the  best-fit  model  is  characterized by  a
stellar mass similar to the mass value predicted by the period spacing
approach,  although  with   a  lower  effective  temperature
(slightly  outside  the  1  $\sigma$  error bar).   However,  at  this
location, and according to  the nonadiabatic pulsation analysis of the
PG1159  sequences employed  here (C\'orsico  et al.   2006)  the model
should be pulsationally stable against $\ell= 1$ modes
and the rates of period changes all positive, in disagreement 
with the observations. In this sense,
we must warn the reader that the location of the theoretical blue edge
of the GW Vir instability strip shown in Fig.
\ref{teff-g} is not to be taken at face value. This is because
changes in  the surface abundances  predicted by stellar  modeling are
expected to  shift the predicted  blue edge location. For instance,
Quirion et al. (2007) have found that increasing the carbon and 
oxygen abundance in  the outer  layers shifts the  blue edge  to  
larger effective temperatures. In particular, by varying the oxygen 
abundance from $X_{\rm O} \sim 0.20$ to $0.40$ 
(with He being replaced by O) and keeping fixed the carbon abundance 
at $X_{\rm C} \sim 0.40$, the blue edge of the GW Vir instability strip
can be pushed to higher effective temperatures by about $10\, 000$ K. 

We  can independently  constrain the  location  of \pg\  in the  $\log
T_{\rm eff}  -\log g$ plane by  examining the trend  of secular period
changes.  In  fact, CK07  have reported  the period
change  of 27 pulsation  modes.  They  found that  some modes  show an
increase in their periods, while  other exhibit a clear decrease. This
could be reflecting that \pg\ is a pre-white dwarf that  still has
significant ongoing contraction. 
To elaborate this point further, we mark in Fig.
\ref{teff-g} with thin dotted lines the limits of the 
region  for  which  our  PG1159  sequences  predict  the  simultaneous
presence of modes with positive and negative $\dot{\Pi}$
\footnote{Above the upper bound of this region, all of the 
modes are expected to have  a decrease in their periods.}.  
To  be consistent  with the  findings of  CK07 our
best-fit model for \pg\ should be located at earlier evolutionary stages, 
as suggested 
by spectroscopy.   In fact, our  predicted best-fit model falls  in the
region of  the $\log  T_{\rm eff}  -\log g$ diagram  where all  of the
pulsational  modes are  expected to  exhibit a  positive value  of the
period change.

Finally, we note that generally the rates of period change measured by
CK07 are about  one order of  magnitude larger than
the values predicted by the  best-fit model. This is particularly true
for the  large amplitude mode  at 517.1 s,  for which CK07 report  a 
value  $\dot{\Pi}= (+18.2  \pm 0.8)  \times 10^{-11}$
s/s, at  odds with our  theoretical value $\dot{\Pi}=  1.40 \times
10^{-11}$ s/s   --- the same  problem occurs with all  other previous
calculations, including those of Kawaler (1986). Also, 
our best-fit model  for \pg\ fails  to reproduce the 
large values of the second order rates of period changes
observed in the star, 
and the contraction and cooling rates estimated by CK07.
Those authors do point out that these estimates need corroboration.
 
The  results of  the period-fit  procedure  carried out  in this  work
suggest  that  the  asteroseismological  mass  of  \pg\  ($\sim  0.565
M_{\odot}$) could  be $\sim 4  \%$ lower than thought  hitherto ($\sim
0.59 M_{\odot}$; see  WEA91, KB94  and more recently CEA07) 
and in closer  agreement with the  spectroscopic mass of
$0.54 M_{\odot}$ as derived by Miller Bertolami \& Althaus (2006) (see
also  Dreizler  \&  Heber  1998). This suggests  that  a  reasonable
consistency  between   the  asteroseismological mass and the spectroscopic
 mass should be  expected  when  the same  evolutionary
tracks are used for both derivations of $M_*$, as we have done 
in C\'orsico et al. (2007c) for \pp\ and we do in the present  
work for \pg.  The  exception to this  assertion is  \rx, for
which we found an  asteroseismological mass about $25 \%$ \emph{lower}
than the  spectroscopic value (C\'orsico et al.  2007ac), 
employing the  same stellar evolution
and pulsation modeling as in the present work.
As we suggested in that paper, the  discrepancy in mass  could be due  
to large errors  in the
spectroscopic  determination of  $\log  g$ and  $T_{\rm  eff}$ for  RX
J2117.1+3412. Uncertainties in the location of the evolutionary tracks
in the HR and $\log T_{\rm eff} - \log g$ diagrams due to the modeling
of PG1159 stars and their precursors  can be discarded in view of the
recent  work by  Miller  Bertolami \&  Althaus  (2007b). They 
conclude that the tracks of PG1159 stars of 
Miller Bertolami \& Althaus (2006) are robust enough as to be used for 
spectroscopical mass determinations of PG 1159-type stars.

In closing, in this paper we have been able to find a PG1159 model
that reproduces the observed pulsation
periods of \pg\ without invoking  
any artificial  adjustments of the structural parameters (the O/C chemical 
profile, the surface He abundance, or the thickness of the He-rich envelope) 
which, instead, are kept fixed  at the values  predicted by  our 
evolutionary  computations. In particular, our PG1159 models are 
characterized by  thick He-rich envelopes. 

However, our best-fit model is unable to explain a number of 
important observed properties of \pg, such as: 

\begin{itemize}

\item the nature itself of \pg\ as a variable star, because  the 
best-fit model lies outside the theoretical dipole GW Vir instability 
domain, i.e., the nonadiabatic treatment indicates pulsational stability
of the model against $\ell= 1$ $g$-modes,
\item the larger range of the period spacings of \pg, as compared 
with those of the best-fit model,   
\item the mixture of positive and negative rates of period change 
measured in \pg,  because the best-fit model has all the theoretical 
rates of period changes positive,
\item the large magnitude of the rates of period change detected  in \pg, 
because they are an order of magnitude larger than the theoretical values 
of the best-fit model.
\end{itemize}

In view of the above shortcomings, we must investigate if
\pg\ could harbor a thinner helium-rich envelope than predicted by our 
evolutionary models, a possibility sustained by  the fact  that PG 1159 
and born-again stars are observed to suffer from appreciable mass loss. 
Further asteroseismological analysis on the basis of PG1159 evolutionary 
models characterized by thin He-rich envelopes, and additional observational 
campaigns will be needed to make definite conclusions about the internal 
structure and evolutionary status of \pg.

  
\begin{acknowledgements}  
The authors would like to  acknowledge the comments and suggestions of
the  referee, Dr. Simon  Jeffery, which  strongly improved  both the
contents  and  presentation of  the  paper.   This  research has  been
partially supported by the PIP  6521 grant from CONICET. This research
has made use of NASA's Astrophysics Data System.
\end{acknowledgements}

\end{document}